\documentclass{JHEP}

\usepackage{amssymb}
\usepackage{epsfig}

\newcommand{\Ree}{\mathop{\rm Re}\nolimits}

\def\rohat{\hat{\rho}}
\def\k{\mathbf{k}}
\def\N{N_{c}}
\def\dt{\mathbf{\Delta}}
\def\V{V_c}
\def\as{\alpha_s}
\def\asb{a}
\def\q{\mathbf{q}}
\def\FF{\mathcal{F}}
\def\f{\frac}
\def\o{\omega}
\def\wp{\omega_{+}}
\def\wm{\omega_{-}}
\def\g{\gamma}
\def\gL{\gamma_s}
\def\G{\Gamma}
\def\qbar{\bar{q}}
\def\half{\frac{1}{2}}
\def\MS{\hbox{$\overline{\mathrm{MS}}$}}
\newcommand{\nn}{\nonumber}
\def\ku{k_1}
\def\kd{k_2}
\newcommand\pu{p_1}
\newcommand\pd{p_2}
\newcommand\pt{p_4}
\newcommand\pq{p_3}

\def\gs{\overline{\g}}
\def\os{\overline{\o}}

\preprint{CERN-TH/2000-361\\
Edinburgh 2000-25 \\
FERMILAB-PUB-00/329-T\\
{\tt hep-ph/0101199}}

\title{Heavy quark production at high energy}

\author{Richard D. Ball\\ Theory Division, CERN,
CH-1211\\ Gen\`eve 23, Switzerland, and\\ Department of Physics
and Astronomy\\ University of Edinburgh, Mayfield Road\\
Edinburgh EH9 3JZ, Scotland\\ Email: \email{richard.ball@cern.ch}}

\author{R. Keith Ellis\\ Theory Department, Fermi National
Accelerator Laboratory,\\ P.O. Box 500, Batavia, IL 60510, USA\\
Email: \email{ellis@fnal.gov}}

\abstract{We report on QCD radiative corrections to heavy quark
production valid at high energy. The formulae presented will
allow a matched calculation of the total cross section which is
correct at $O(\as^3)$ and includes resummation of all terms of
order $\as^3 [\as \ln (s/m^2)]^n$. We also include asymptotic
estimates of the effect of the high energy resummation. A
complete description of the calculation of the heavy quark impact
factor is included in an appendix.}


\begin{document}

\section{Introduction}

In this paper we report on the calculation of the strong
radiative corrections to the~ process
\begin{equation}
p + \bar{p} \longrightarrow b + \bar{b}+X\,.
\end{equation}
This reconsideration is prompted in part by the fact that
measurements at the \mbox{Fermilab}
Tevatron~\cite{ref:DO_bx}--\cite{Abbott:2000wu} indicate that the
$b$ cross section lies above the
predictions~\cite{Nason:1988xz,Mangano:1992jk} from $O(\as^3)$
perturbation theory~\cite{Nason:1988xz,Beenakker:1991ma}. Similar
excesses have been reported recently in the photoproduction $b$
cross-section~\cite{Adloff:1999nr,Breitweg:2000nz}, and in $b$
production in $\g\g$
collisions~\cite{Acciarri:2000kd,Csilling:2000xk}. Two remarks
need to be made about this situation. First, the transformation
of the observed experimental cross section to a $b$-quark cross
section requires the inclusion of fragmentation and decay
corrections which are subject to theoretical errors. Although
these corrections may be smaller when we consider the $b$-jet
cross section~\cite{Frixione:1997nh} rather than the $b$-quark
cross section, experimental results for the $b$-jet cross section
still indicate an excess over theoretical
predictions~\cite{Abbott:2000iv}. Second, the fixed order
perturbation theory description of $b$-quark production receives
large corrections in order $\as^3$ which call into question the
validity of perturbation theory. We may simply conclude that, in
this energy and mass range, the $\as^3$ theory is not very
predictive and ignore the disagreement with experiment as an
inadequacy of the theory. A more constructive approach, which we
adopt in this paper, is to try and resum the large corrections.

The heavy quark cross section is calculable because the heavy
quark mass $M$ is larger than the QCD scale $\Lambda$. As first
observed in ref.~\cite{Nason:1988xz} the large corrections in
$O(\as^3)$ come both from the threshold region $s \sim M^2$ and
from the the high energy region $s \gg M^2$, where $s$ is the
normal Mandelstam variable for the parton \mbox{sub-process.}

The resummation of the threshold soft gluon corrections is best
carried out in $\o$-space, where $\o$ is the moment variable after
Mellin transform with respect to $s$. In $\o$-space the threshold
region corresponds to the limit $\o \rightarrow \infty$ and the
structure of the threshold corrections is as follows
\begin{equation}
\hat{\sigma} = \sigma^{LO} \left\{ 1 + \sum_{j=1}^{\infty}
\sum_{k=1}^{2 j} b_{jk} \alpha_s^j \ln^k \o \right\}.
\end{equation}
Detailed studies~\cite{Bonciani:1998vc} indicate that such
threshold resummation is of limited importance at the Tevatron
where $b$-quarks are normally produced far from threshold.
Resummation effects lead to minor changes in the predicted cross
section and only a slight reduction in the scale dependence of
the results.

At high energy, heavy quark production becomes a two scale
problem, since $S \gg M^2 \gg \Lambda^2$ and the short distance
cross section contains logarithms of $s/M^2$. As we proceed to
higher energies such terms can only become more important, so an
investment in understanding these terms at the Tevatron will
certainly bear fruit at the LHC. At small $\o$ the dominant terms
in the heavy quark production short distance cross section are of
the form
\begin{equation}\label{logs}
\hat{\sigma}\sim \f{\as^2}{M^2} \sum_{j=0}^\infty \as^{j}
\sum_{k=0}^{\infty} c_{jk} \left(\f{\as}{\o} \right)^{k}.
\end{equation}
For $j=0$ we obtain the leading logarithm series (LLx).
Resummation of these terms has received considerable theoretical
attention in the context of the $k_T$-factorization
formalism~\cite{Ellis:1990hw}--\cite{Camici:1997ta}. Although
some work quantitative nature has been
performed~\cite{Hagler:2000dd,Ryskin:2000wz}, so far the
connection with the precise structure function data has been
indirect since the unintegrated gluon distributions are not
measured directly. In addition, both the difficulty in
reconciling $k_T$-factorization with structure function data from
HERA~\cite{Ball:1994du}--\cite{Blumlein:1998em}, and the
theoretical uncertainty in the face of large~\cite{Fadin:1998py}
destabilizing subleading
corrections~\cite{Ross:1998xw}--\cite{Armesto:1998gt} have
hampered this program. However significant progress in
understanding the origin and resolution of these instabilities
has been made
recently~\cite{Salam:1998tj}--\cite{Altarelli:2000vw}, and
resummed parton densities in the small $x$ region can be
determined from HERA data~\cite{Altarelli:2000mh}. Consistent
resummed calculations of heavy quark production cross-sections are
thus finally a possibility.

The aim of this paper is to collect together the relevant results
for a quantitative calculation of the total heavy quark
production cross section. Our formulae include the known fixed
order results, as well as the resummation of the leading tower of
high energy logarithms. The approach to this resummation is as
follows. The $O(\as^2)$ and $O(\as^3)$ total cross sections are
well known at the parton level. Using the results of the
$k_T$-factorization, valid at high energy, we can resum the
logarithms of $s/M^2$ by determining the leading coefficients
$c_{0k}$ of eq.~(\ref{logs}) for $k=2,\dots,\infty$, resumming
the singularities using the techniques of
ref.~\cite{Altarelli:2000vw}, and then combining with resummed
parton distributions~\cite{Altarelli:2000mh} to give a high-energy
improved short distance cross section.

This paper deals exclusively with results for the total cross
section. This quantity is not experimentally accessible with
current detectors which typically have a $p_T$ threshold below
which $b$-quarks cannot be observed. The technical problems of
extending this analysis to less inclusive measurements may be
formidable. The present analysis is intended to give us an
understanding of the size of these terms, before undertaking the
extension to experimentally more realistic quantities.

The structure of this paper is as follows. In
section~\ref{sec:2}, we consider the description of the direct
component of bottom quark photoproduction both in fixed order
perturbation theory and in the framework of $k_T$-factorization.
We use the latter formalism as a technical device to resum the
logarithms of $s/M^2$ and cast the final result in the form of a
high-energy improved collinear factorization formula.
Section~\ref{sec:3} repeats similar steps for the hadroproduction
of heavy flavours. In practice, the treatment of the
photoproduction of heavy flavours is more complex than described
in section~\ref{sec:2} because of the hadronic (resolved)
component of the photon. The resolved component of the
photoproduction cross section should be treated by the methods of
section~\ref{sec:3}. Similar remarks apply to heavy quark
production in $\g\g$ collisions. Section~\ref{sec:4}
presents some analytic estimates. In section~\ref{sec:5} we draw
some preliminary conclusions. A complete description of the
calculation of the heavy flavour impact factors is presented in
appendix~\ref{appa}.

\section{Photoproduction of heavy flavours}\label{sec:2}

\subsection{Fixed order perturbation theory}

In this section we report on the results for the direct component
of the photoproduction of heavy flavours in fixed order
perturbation theory~\cite{Ellis:1989sb}. We first define reduced
cross sections $\Sigma$ with the mass dimensions removed, both at
the hadronic and partonic level,
\begin{equation}
\Sigma(\rho)=M^2 \sigma(M^2,S)\,,\qquad
\hat{\Sigma}_{\gamma j}\left(\rohat,\f{\mu^2}{M^2}\right)=
M^2 \hat{\sigma}_{\gamma j}(s,M^2,\mu^2) \,.
\end{equation}
Here $S$ and $s$ are the squares of the total hadronic and
partonic centre of mass energies, respectively, $M$ the mass of
the heavy quark, $\rho=4 M^2/S,\rohat=4 M^2/s=\rho/x$ and $\mu$ is
the renormalization and factorization scale. We can further
express the partonic cross section as
\begin{equation}
\hat{\Sigma}_{\gamma j} \left(\rohat,\f{\mu^2}{M^2}\right) =
\alpha_{em} \alpha_s(\mu^2) f_{\gamma j}
\left(\rohat,\f{\mu^2}{M^2}\right).
\end{equation}
The photon-hadron cross section is then given by the collinear
factorization formula
\begin{equation} \label{gamtotsec}
\Sigma (\rho) = \alpha_{em} \alpha_s(\mu^2) \int_\rho^1
\frac{dx}{x} \sum_{j=q,\qbar,g} f_{\gamma j} \left(\frac{\rho}{x},
\f{\mu^2}{M^2}\right) F^j(x,\mu^2) \,,
\end{equation}
where $F^j(x,\mu^2)=x q(x,\mu^2),x\qbar (x,\mu^2),xg(x,\mu^2)$,
$j=q,\qbar ,g$ are the parton momentum densities, renormalized in
\MS\ at scale $\mu$. The coefficient functions $f_{\gamma j}$
have a perturbative expansion which, following
ref.~\cite{Ellis:1989sb}, we write as,
\begin{equation}
f_{\gamma j}\left(\rho,\f{\mu^2}{M^2}\right) =
f^{(0)}_{\gamma j}(\rho) + 4 \pi \alpha_s(\mu^2)
\left[f^{(1)}_{\gamma j}(\rho) + \bar{f}^{(1)}_{\gamma j}(\rho)
\ln \f{\mu^2}{M^2} \right]+ O(\as^2)\,.
\end{equation}
The functions $f_{\gamma j}$ depend on the charge of the quark
which interacts with the photon. To make these dependences
explicit we further define the quantities,
\begin{eqnarray}
f_{\gamma g}\left(\rho,\f{\mu^2}{M^2}\right) &=& e_{Q}^2
c_{\gamma g}\left(\rho,\f{\mu^2}{M^2}\right),\nn \\
f_{\gamma q}\left(\rho,\f{\mu^2}{M^2}\right) &=&
f_{\gamma \qbar}\left(\rho,\f{\mu^2}{M^2}\right)
= e_{Q}^2 c_{\gamma q}\left(\rho,\f{\mu^2}{M^2}\right)
+ e_{q}^2 d_{\gamma q}\left(\rho,\f{\mu^2}{M^2}\right),
\end{eqnarray}
where $e_q$ is the charge of the incoming light quark, while
$e_Q$ is the charge of the heavy quark. Results for the
perturbative expansions of the functions, $c_{\gamma g},
c_{\gamma q}$ and $d_{\gamma g}$ are given in
ref.~\cite{Ellis:1989sb}.

We can also define the moments $f_{\o}$ of all functions $f$,
\begin{equation} \label{omegamoments}
f_\o = \int_0^1 d \rho \; \rho^{\o-1} f(\rho)\,,\qquad
F_{\o}(\mu^2 )= \int_0^1 d x \; x^{\o-1} F(x,\mu^2)\,.
\end{equation}
Taking moments of eq.~(\ref{gamtotsec}) we get
\begin{equation}
\Sigma_{\o}= \int_0^1 d\rho \; \rho^{\o-1} \Sigma(\rho) =
\alpha_{em} \alpha_s(\mu^2) \sum_{j=q,\qbar,g} f_{\gamma j \; \o}
\left(\f{\mu^2}{M^2}\right) F^{j}_{\o}(\mu^2)\,.
\end{equation}
The moments of $c_{\gamma g},c_{\gamma q}$ and $d_{\gamma g}$ are
easily obtained from the results of~\cite{Ellis:1989sb}. The
expressions for the partonic cross-sections $\hat{\Sigma}$ are
then
\begin{eqnarray}
\hat{\Sigma}^{\gamma g}_\o \left(\f{\mu^2}{M^2}\right)&=&
e_Q^2 \alpha_{em} \alpha_s \left(c_{\gamma g \; \o }^{(0)} +
4\pi\as \left[c_{\gamma g \; \o }^{(1)} + \ln
\left(\f{\mu^2}{M^2}\right) \bar{c}_{\gamma g \; \o }^{(1)}
\right]\right),\nn\\
\hat{\Sigma}^{\gamma q}_\o\left(\f{\mu^2}{M^2}\right)&=&
\alpha_{em} \alpha_s^2 4\pi \left(e_q^2 \left[d_{\gamma q \;
\o }^{(1)}+\ln\left(\f{\mu^2}{M^2}\right)\bar{d}_{\gamma q \;
\o }^{(1)}\right]\right.+\nn\\
&&\hphantom{\alpha_{em} \alpha_s^2 4\pi \Bigg(}\!\left. +\,e_Q^2
\left[c_{\gamma q \; \o }^{(1)} + \ln\left(\f{\mu^2}{M^2}\right)
\bar{c}_{\gamma q \; \o}^{(1)} \right]\right).\label{nlopartxsecg}
\end{eqnarray}
The form of these results in the high energy limit are then
readily derived~\cite{Ellis:1990hw}: as $\o\rightarrow 0$
\begin{eqnarray}
\hat{\Sigma}^{\gamma g}_{\o}\left(\f{\mu^2}{M^2}\right) &\sim&
\pi e_Q^2 \alpha_s \alpha_{em} \left[\frac{7}{9} +
\left(\frac{41}{27} -\frac{7}{9} \ln \frac{\mu^2}{M^2}\right)
\frac{\asb}{\o} \right] +\cdots\,,\nn\\
\hat{\Sigma}^{\gamma q}_{\o}\left(\f{\mu^2}{M^2}\right) &\sim&
\pi e_Q^2 \alpha_s \alpha_{em}\frac{\V}{2 \N^2}
\left(\frac{41}{27}- \frac{7}{9}\ln\frac{\mu^2}{M^2}
\right)\frac{\asb}{\o} +\cdots \,,\label{photnlosingg}
\end{eqnarray}
where as usual $\asb =\N \alpha_S/\pi$, and $\V=\N^2-1$. The
$d_{\gamma g}$ terms are non singlet, and thus contain no $\o=0$
singularities.

\subsection{High energy behaviour}

We now turn to the high energy behaviour of the photoproduction
cross sections which contain logarithms of $s/M^2$ and need to be
resummed. This resummation is performed by using the
${\k}$-factorized expression for the cross section
\begin{equation} \label{gamktfac}
\Sigma(\rho) =\sum_{j=q,\qbar,g}\int \; \f{dx}{x}\;
\frac{d^2{\k}}{\pi} \; \hat{\Sigma}_{\gamma j}
\left(\frac{\rho}{x},{\k},M^2,\mu^2\right)\FF^j(x,{\k}^2,\mu^2)\,,
\end{equation}
The functions $\FF^j$ are the unintegrated parton distribution
functions and $\hat{\Sigma}_{\gamma j}$ are the (lowest-order and
gauge-invariant) off-shell continuations of the parton cross
sections. The diagrams necessary to calculate $\hat{\Sigma}$ are
illustrated in figure~\ref{photoprod} where $k_2=x p_2 + {\k}$.
We can simplify eq.~(\ref{gamktfac}) by taking the moments with
respect to the longitudinal variable, cf.\
eq.~({\ref{omegamoments}}), to undo the convolution:
\FIGURE[t]{\epsfig{file=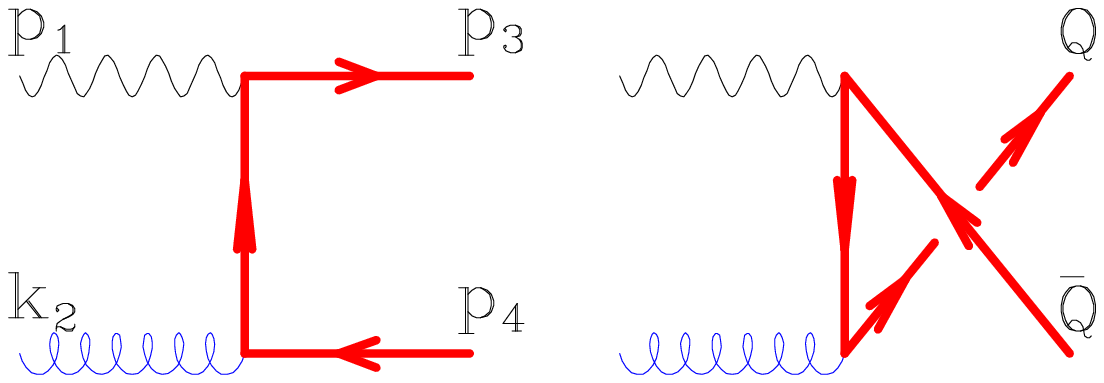,width=.6\textwidth}
\caption{Diagrams for the direct component of heavy quark
photoproduction at high energy.}\label{photoprod}}
\begin{equation} \label{gamktfacmel}
\Sigma_\o = \sum_{j=q,\qbar,g} \int \frac{d^2{{\k}}}{\pi} \;
J^j_\o\left(\f{{\k}^2}{M^2},\f{{\mu}^2}{M^2}\right)
\mathcal{F}^j_\o({\k}^2;\mu)\,.
\end{equation}
$J^j_\o$ is the impact factor which expresses the response of the
quark-antiquark pair as a function of the transverse momentum of
the incoming light parton. A further simplification can be
achieved by defining Mellin transforms with respect to the
transverse momentum squared variables:
\begin{eqnarray}
\FF_{\o}(\g)&=&\int_{-\infty}^\infty \f{d^2{\k}}{\pi}
\left(\f{{\k}^2}{\mu^2}\right)^{-\g}\FF_{\o}({\k}^2,\mu^2)\,,\nn\\
J_\o(\gamma) &=& \int_{-\infty}^\infty \frac{d^2 {\k}}{{\pi \k}^2}
\left(\frac{{\k}^2}{M^2}\right)^\gamma J_\o\left(\f{{\k}^2}{M^2},
\f{{\mu}^2}{M^2}\right),\label{melunintpdf}
\end{eqnarray}
where the dependence of the Mellin transforms on $\mu^2$ has now
been suppressed. The Mellin transform of $\FF$ is defined on the
fundamental strip $\alpha< \Ree(\g)<\beta$ corresponding to
the assumed behaviours
\begin{eqnarray}
\FF_{\o}({\k}^2,\mu^2) &\sim& ({\k}^2)^{\alpha-1}\,,
\qquad{\k}^2 \rightarrow\infty\,,\nn \\
\FF_{\o}({\k}^2,\mu^2) &\sim& ({\k}^2)^{\beta-1}\,,
\qquad {\k}^2 \rightarrow 0 \nn \,.
\end{eqnarray}

The existence of the integrated parton distributions
\begin{equation} \label{intpdf}
F_{\o}(\mu^2)= \int_0^{\mu^2} d{\k}^2 \FF_{\o }({\k}^2,\mu^2)
\end{equation}
implies that $\beta>0$. For the impact factor $J$ the
Mellin transform is defined in the strip $0<\mbox{Re}(\g)<1$.
The inverse Mellin transforms are defined as,
\begin{eqnarray}
\FF_{\o}({\k}^2,\mu^2) &=& \frac{1}{{\k}^2} \int_C
\frac{d \gamma}{2 \pi i} \left(\frac{\k^2}{\mu^2}\right)^{\gamma}
\FF_{\o}(\gamma)\,,\nn\\
J_{\o}\left(\f{{\k}^2}{M^2},\f{{\mu}^2}{M^2}\right) &=&
\int_C \frac{d \gamma }{2 \pi i } \left(\frac{M^2}{\k^2}
\right)^{\gamma} J_{\o}(\gamma)\,,\label{invmel}
\end{eqnarray}
where $C$ is a contour lying in the fundamental strips defined
above. Inserting these expressions in the representation,
eq.~(\ref{gamktfacmel}), and performing the transverse momentum
integration we then have
\begin{equation} \label{gamktfacmelmel}
\Sigma_{\o}= \sum_{k=q,\bar{q},g} \int_C\; \frac{d\gamma}{2 \pi i}
\left(\frac{M^2}{\mu^2} \right)^\gamma J^k_\o(\gamma)\;
\FF^k_\o(\g) \,.
\end{equation}

To understand how this result may be used to resum high energy
logarithms, consider first the unintegrated gluon density
$\FF^g_{\o}({\k}^2,\mu^2)$ as a solution to the BFKL equation:
the Mellin transform $\FF^g_{\o}(\g)$ then has the form
\begin{equation}\label{lippole}
\FF^g_{\o}(\g) = \frac{\FF^{(0)}_\o(\g)}{1-\frac{\asb}{\o}
\chi_0(\g)}\,,
\end{equation}
where $\chi_0(\g)= 2 \psi(1) -\psi(\g) -\psi(1-\g)$ is the Mellin
transform of the lowest order BFKL kernel. Assuming that the
asymptotic behaviour $(\k^2 \gg \mu^2)$ is determined
perturbatively (rather than by the properties of $\FF^{(0)}_\o$),
the rightmost singularity of $\FF^g_{\o}(\g)$ is then a simple
pole in the fundamental strip at $\gamma = \gL(\asb/\o)$, where
$\gL(\asb/\o)$ is defined implicitly through the relation
\begin{equation}\label{duality}
\f{\asb}{\o}\chi_0\left(\gL\left(\f{\asb}{\o}\right)\right)= 1\,,
\end{equation}
and the condition $\gL(0)=0$. When expanded perturbatively we have
\begin{equation}\label{lippert}
\gL\left(\f{\asb}{\o}\right)=\frac{\asb}{\o} +2\zeta_3
\left(\frac{\asb}{\o}\right)^4 +2\zeta_5
\left(\frac{\asb}{\o}\right)^6 + O\left(\left(\frac{\asb}{\o}
\right)^7\right).
\end{equation}
It follows that in the approximation in which we keep only the
residue of this simple pole, the inverse Mellin transform,
eq.~(\ref{invmel}) is given by
\begin{equation} \label{Fdef}
\FF^g_{\o}({\k}^2,\mu^2) = -\frac{1}{{\k}^2} \frac{\chi_0
(\gL(\asb/\o))}{\chi_0^\prime (\gL(\asb/\o))}
\left(\frac{\k^2}{\mu^2}\right)^{\gL(\asb/\o)} \FF^{(0)}_\o
\left(\gL\left(\f{\asb}{\o}\right)\right).
\end{equation}
In obtaining eq.~(\ref{Fdef}) we have taken the limit ${\k^2} >
\mu^2$ so that we can neglect the contributions from
non-perturbative singularities in the input distribution
$\FF^{(0)}_\o(\g)$, as well as higher twist perturbative
contributions coming from solutions to eq.~(\ref{duality}) to the
left of $\gL(\asb/\o)$). Substituting eq.~(\ref{Fdef}) into the
definition eq.~(\ref{intpdf}) of the integrated gluon
distribution we get
\begin{equation}\label{intglu}
F^g_\o(\mu^2)=-\frac{1}{\gL(\asb/\o)}\frac{\chi_0(\gL(\asb/\o))}
{\chi_0^\prime (\gL(\asb/\o)))} \FF^{(0)}_{\o}
\left(\gL\left(\frac{\asb}{\o}\right)\right).
\end{equation}
High energy factorization makes a definite prediction for the
evolution of the integrated gluon distribution in terms of the
function $\g_s$:
\begin{equation}\label{intgluevol}
F^g_\o(M^2)= \left(\frac{M^2}{\mu^2}\right)^{\gL(\asb/\o)}
F^g_\o(\mu^2)\,.
\end{equation}

In a similar way, if the integral eq.~(\ref{gamktfacmelmel}) may
also be dominated by the simple pole at
$\gamma = \gL(\asb/\o)$ (because the impact
factor is regular in the fundamental strip), we find
that the gluonic contribution to $\Sigma_{\o}$ is
\begin{eqnarray}
\Sigma_{\o}&=& \left(\frac{M^2}{\mu^2}\right)^{\gL(\asb/\o)}
J^g_\o \left(\gL\left(\f{\asb}{\o}\right)\right)
\frac{\chi_0 \left(\gL(\asb/\o)\right)}{\chi_0^\prime (\gL)}
\FF^{(0)}_\o\left(\gL\left(\f{\asb}{\o}\right) \right)+\cdots\nn\\
&=& \gL\left(\f{\asb}{\o}\right) J^g_\o \left(\gL
\left(\f{\asb}{\o}\right)\right)\left(\frac{M^2}{\mu^2}
\right)^{\gL(\asb/\o)} F^g_\o(\mu^2) +\cdots\nn\\
&=& j\left(\gL\left(\f{\asb}{\o}\right)\right)
\left(\frac{M^2}{\mu^2}\right)^{\gL(\asb/\o)}
F^g_\o(\mu^2) +\cdots\,. \label{gamcollfacsingglu}
\end{eqnarray}
where in the second line we substituted eq.~(\ref{intglu}) for
the integrated gluon distribution, and in the last used the fact
that at high energy (i.e. $\o\rightarrow 0$) we may~write
\begin{equation}\label{Jjdef}
\gamma J^g_\o(\gamma) = j(\gamma) + O(\o)\,,
\end{equation}
since the impact factor $J_\o(\gamma)$ is free of singularities
and terms of $O(\o)$ are subleading.

\pagebreak[3]

This simple argument is however not sufficient to define the high
energy cross-section, since if we redefine the unintegrated gluon
distribution through a scheme change $\FF^g_{\o}(\g)\rightarrow
u(\g)\FF^g_{\o}(\g)$, the function
$\chi(\g,\as)=\chi_0(\g)+\as\chi_1(\as)+\cdots$ only changes at
NLLx~\cite{Catani:1996ze,Ball:1995tn}:
\begin{equation}\label{scheme}
\chi(\g)\longrightarrow\chi(\g) - \as b_0\chi(\g)
\frac{d}{d\gamma}\ln u(\gamma)+O(\as^2)\,.
\end{equation}
where, as usual, $b_0=\f{11\N -2 n_f}{12 \pi}$. Consequently
unless the BFKL equation is treated at NLLx, the effect of the
impact factor can always be removed by a scheme change.
Fortunately the NLLx kernel $\chi_1(\g)$ is now
known~\cite{Fadin:1998py}, and techniques to stabilise the
expansion have now been developed which make meaningful NLLx
resummations
possible~\cite{Salam:1998tj}--\cite{Altarelli:2000mh}. All that
remains is to specify the particular factorization scheme in
which the impact factor is to be calculated. Impact factor
calculations such as those described in appendix~\ref{appa},
based on the evaluation of off-shell amplitudes implicitly employ
a ``$k_T$-factorization scheme''. For the photoproduction of heavy
quarks, the result of such a calculation
is~\cite{Ellis:1990hw,Catani:1990xk}, cf.\
eq.~(\ref{onelegonshell})
\begin{equation} \label{jdef}
j(\gamma) = e_Q^2 \alpha_{em} \alpha_s\; \frac{\pi}{3}
\frac{7-5 \gamma}{3-2 \gamma} B(1-\gamma,1-\gamma)
B(1+\gamma,1-\gamma)\,.
\end{equation}
We note parenthetically that the lowest order expression for
$J^g_\o(\gamma)$ including the full $\o$ dependence has been
given in ref.~\cite{Catani:1992zc}. In our notation it is
\begin{eqnarray}
J^g_\o(\gamma) &=& e_Q^2 \alpha_{em} \alpha_s \; 4^{\o} \pi \;
B(1-\gamma+\o,1-\gamma+\o) B(1+\gamma,1-\gamma+\o) \times
\nonumber \\&& 
\times \,\frac{[14 +20 \o+9 \o^2+\o^3-\gamma (10+7 \o+\o^2)]}
{\gamma (3+\o) (2+\o) ( 3-2 \gamma+2 \o)}
\end{eqnarray}

For small $\gamma$
\begin{equation} \label{jexp}
j(\gamma)\sim e_Q^2 \alpha_{em} \alpha_s \left[\frac{7\pi}{9} +
\frac{41\pi }{27} \gamma + O(\gamma^2)\right].
\end{equation}
This can be related to the result in \MS\ factorization through a
scheme change $u(\g)=R(\g)$, where $R(\g)$ is a universal
function which fixes the normalization of the \MS\ gluon
distribution~\cite{Catani:1994sq}. Explicitly
\begin{eqnarray}
R(\g) &=& \left[\frac{\Gamma(1-\g) \chi(\g)}
{(\Gamma(1+\g) (-\g) \chi^\prime (\g))}\right]^{1/2}\times\nn\\
&&\times\,\exp \left[\g \psi(1)+ \int_0^{\g} \; dx
\frac{(\psi^\prime(1) -\psi^\prime(1-x))}{\chi(x)}\right],
\label{Rdefinition}
\end{eqnarray}
so for small $\g$
\begin{equation}\label{Rexp}
R(\g) = 1 +\f{8}{3}\zeta_3\g^3 -\f{3}{4}\zeta_4\g^4
+\f{22}{5}\zeta_5\g^5 +O(\g^6)\,,
\end{equation}
where $\zeta_n$ is the Riemann zeta function. In \MS\
factorization we thus have
\begin{equation}\label{photcomsg}
\gamma J^g_\o(\gamma) = j(\gamma)R(\gamma) + O(\o)\,,
\end{equation}
Note that once we have fixed the factorization scheme, the
leading order calculation of the impact factor is sufficient for
a consistent NLLx calculation: if we knew the impact factor at
NLO, we would also need the function $\chi$ at NNLLx for a fully
consistent calculation.

It remains to add in the quark contribution. At LLx $\g_{gq}\sim
r_c \g_{gg}$, where $r_c$ is defined as the ratio of colour
charges:
\begin{equation}
r_c = \frac{\V}{2\N^2} = \f{C_F}{C_A} = \f{4}{9}
\end{equation}
when $\N=3$. It follows that a quark may turn into a gluon
provided one includes an additional factor of $r_c$. However
$\g_{qg}$ is NLLx, so if the gluon turns back to a quark it costs
an additional power of $\as$. Consequently at LLx the (bare)
unintegrated quark and gluon distributions are related by
\begin{equation}
\FF^q_{\o}({\k}^2,\mu^2) = r_c\left(\FF^g_{\o}({\k}^2,\mu^2)
-\delta^{(2)}(\k)\right).
\end{equation}
It is then straightforward to prove, along the lines of similar
arguments presented in ref.~\cite{Catani:1994sq}, that at LLx the
impact factor for quarks is related to that for gluons:
specifically in \MS\
\begin{equation}\label{photcomsq}
\g J^q_\o(\gamma) = \g J^{\qbar}_\o(\gamma)
=r_c(j(\gamma)R(\gamma) -j(0)) + O(\o)\,.
\end{equation}

Putting all these results together, we find that at high energies
the partonic cross-sections in \MS\ factorization are given by
\begin{eqnarray}
\hat{\Sigma}^{\gamma g}_\o &=& j\left(\gL\left(\f{\asb}{\o}\right)
\right)R\left(\gL\left(\f{\asb}{\o}\right)\right)\left(\frac{M^2}
{\mu^2} \right)^{\gL(\asb/\o)}\,,\nn\\
\hat{\Sigma}^{\gamma q}_\o &=& \hat{\Sigma}^{\gamma \qbar}_\o =
r_c \left[j\left(\gL\left(\f{\asb}{\o}\right)\right)R\left(\gL
\left(\f{\asb}{\o}\right)\right)\left(\frac{M^2}{\mu^2}
\right)^{\gL(\asb/\o)}-j(0)\right].\label{xsecMSsingq}
\end{eqnarray}

\subsection{Double leading resummation}

Using the expansions eqs.~(\ref{lippert}),~(\ref{jexp})
and~(\ref{Rexp})
\begin{equation}
j\left(\gL\left(\f{\asb}{\o}\right)\right)
R\left(\gL\left(\f{\asb}{\o}\right)\right) \sim e_Q^2
\alpha_{em} \alpha_s\; \left[\frac{7\pi}{9} +\frac{41
\pi}{27}\frac{\asb}{\o}
+O\left(\left(\frac{\asb}{\o}\right)^2\right)\right].
\end{equation}
It follows that the high energy results~(\ref{xsecMSsingq}) are
consistent with the high energy expansions~(\ref{photnlosingg})
of the fixed order results~(\ref{nlopartxsecg}) in the region in
which they overlap. Consequently we may construct ``double
leading'' expansions of the high energy cross-sections by
combining the two, provided we take care to subtract the
terms~(\ref{photnlosingg}) to avoid double counting. For the case
of the gluonic contribution to photoproduction such a matched
calculation has been already presented in
ref.~\cite{Catani:1992zc}.

\looseness=1
So overall the result for the high energy cross-section in double
leading expansion~is
\begin{eqnarray}
\hat{\Sigma}^{\gamma g}_\o &=& e_Q^2 \alpha_{em} \alpha_s
\left\{c_{\gamma g \; \o }^{(0)} + 4\pi\as \left[
c_{\gamma g \; \o }^{(1)} + \ln\left(\f{\mu^2}{M^2}\right)
\bar{c}_{\gamma g\;\o }^{(1)}\right] \right\}+\nonumber \\[2pt]
&& +\,\left[j\left(\gL\left(\f{\asb}{\o}\right)\right)R
\left(\gL\left(\f{\asb}{\o}\right)\right)
\left(\frac{M^2}{\mu^2} \right)^{\gL(\asb/\o)}
-j(0)\left(1+\frac{\asb}{\o} \ln\frac{M^2} {\mu^2}\right)
-\frac{\asb}{\o} j^\prime (0)\right],\nn\\
\hat{\Sigma}^{\gamma q}_\o &=& \hat{\Sigma}^{\gamma \qbar}_\o=
e_q^2 \alpha_{em} \alpha_s^2 4\pi
\left[d_{\gamma q \; \o }^{(1)}+\ln\left(\f{\mu^2}{M^2}\right)
\bar{d}_{\gamma q \; \o }^{(1)}\right]+\nonumber \\[2pt]
&&\hphantom{\hat{\Sigma}^{\gamma \qbar}_\o=}\!+\, e_Q^2
\alpha_{em}\alpha_s 4\pi\as \left[c_{\gamma q \;
\o}^{(1)} + \ln\left(\f{\mu^2}{M^2}\right)
\bar{c}_{\gamma q \; \o}^{(1)}\right]+\nonumber \\[2pt]
&&\hphantom{\hat{\Sigma}^{\gamma \qbar}_\o=}\!+\, r_c
\left[j\left(\gL\left(\f{\asb}{\o}\right)\right)
R\left(\gL\left(\f{\asb}{\o}\right)\right)
\left(\frac{M^2}{\mu^2} \right)^{\gL(\asb/\o)}\right.-\nn\\[2pt]
&&\hphantom{\hphantom{\hat{\Sigma}^{\gamma \qbar}_\o=}\!+\, r_c
\Bigg[}\!\left.-\,j(0)\left(1+\frac{\asb}{\o}
\ln\frac{M^2} {\mu^2}\right)- \frac{\asb}{\o} j^\prime (0)\right],
\end{eqnarray}
where $j(0)=e_Q^2 \alpha_{em} \alpha_s 7\pi/9$ and $j'(0)=e_Q^2
\alpha_{em} \alpha_s 41\pi/27$. The final step is then to
implement the NLLx resummation by replacing $\gL(\asb/\o)$ with a
subtracted anomalous dimension, as described
in~\cite{Altarelli:2000mh}. There is then an ambiguity in the
treatment of the double counting terms, just as in the resummed
double leading expansion of the anomalous dimension: we can
choose whether to resum the full series of singularities
eqs.~(\ref{xsecMSsingq}), or whether
instead to omit the double counting terms from the resummation,
so as to leave the $O(\as^3)$ terms unchanged. This cannot be
resolved at NLLx, and will thus be useful as an estimate of the
overall uncertainty in the~procedure.

\section{Hadronic production of heavy flavours}\label{sec:3}

\subsection{Fixed order perturbation theory}

Before considering high energy resummation we first describe the
results in collinear factorization for the total cross section
for the hadroproduction of heavy quarks of mass $M$. The hadronic
cross section is factorized as
\begin{equation} \label{totsec}
M^2 \sigma(M^2,s) = \Sigma(\rho)= \sum_{i,j} \int \frac{dx_1}{x_1}
\frac{dx_2}{x_2} \hat{\Sigma}_{ij}\left(\frac{\rho}{(x_1x_2)},
\f{\mu^2}{M^2}\right) F^i(x_1,\mu^2) F^j(x_2,\mu^2) \,,
\end{equation}
where the sum runs over all $i,j=q,\qbar,g$,
$\hat{\Sigma}_{ij}=\hat{\Sigma}_{ji}$ and $\rho=4 M^2/ S$.
Following ref.~\cite{Nason:1988xz} the partonic cross section is
given by
\begin{equation} \label{reducedpartxsec}
M^2 \hat{\sigma}_{ij}(s,M^2,\mu^2) \equiv \hat{\Sigma}_{ij}
\left(\rohat,\f{\mu^2}{M^2}\right) = \alpha_s^2(\mu^2) \;
f_{ij}\left(\rohat,\f{\mu^2}{M^2}\right),
\end{equation}
where $\rohat=4 M^2/ s =\rho/(x_1 x_2)$, $s$ being the partonic
centre-of-mass energy. The functions $f_{ij}=f_{ji}$ have a
perturbative expansion
\begin{equation}
f_{ij}\left(\rho,\frac{\mu^2}{M^2}\right)= f^{(0)}_{ij}(\rho) +
4 \pi \alpha_s(\mu^2) \left[ f^{(1)}_{ij}(\rho) + \bar{f}^{(1)}_{ij}
(\rho) \ln \frac{\mu^2}{M^2}\right]+ O(\alpha_S^2)\,.
\end{equation}
The functions $f$ have been calculated in perturbation theory to
NLO in ref.~\cite{Nason:1988xz}, $f^{(0)}$, $\bar{f}^{(1)}$
exactly and ${f}^{(1)}$ as a numerical fit. Note that while
$\hat{\Sigma}_{gg}$ and $\hat{\Sigma}_{q\qbar}
=\hat{\Sigma}_{\qbar q}$ begin at $O(\as^2)$,
$\hat{\Sigma}_{gq}=\hat{\Sigma}_{g\qbar}$ only start at
$O(\as^3)$ and $\hat{\Sigma}_{qq}=\hat{\Sigma}_{\qbar\qbar}$
(currently unknown) start at $O(\as^4)$. Moments $f_{\o}$ of all
functions $f$ may be defined as in eq.~(\ref{omegamoments}).
Taking moments of eq.~(\ref{totsec}) we get
\begin{equation}\label{defnomegamel}
\Sigma_{\o}= \int_0^1 d\rho \; \rho^{\o-1} \Sigma(\rho) =
\alpha_s^2(\mu^2)\sum_{ij} f_{ij\; \o}\left(\frac{\mu^2}{M^2}\right)
F^i_\o(\mu^2) F^j_\o(\mu^2) \,.
\end{equation}
With this normalization, we find the following
results~\cite{Ellis:1990hw} in the high energy limit as
$\o\rightarrow 0$
\begin{eqnarray}
f^{(0)}_{gg\;\o}&\sim& \frac{\pi}{\V} \left(\frac{4}{15}\N -
\frac{7}{18} \frac{1}{\N}\right) +O(\o)\,,\nn \\
f^{(1)}_{gg\;\o}&\sim& \frac{\N}{2 \pi \V } \left(\frac{154}{225}
\N - \frac{41}{54} \frac{1}{\N}\right) \frac{1}{\o} +O(1)\,,\nn \\
\bar{f}^{(1)}_{gg\;\o} &\sim& - \frac{\N}{2\pi \V}
\left(\frac{4}{15} \N -\frac{7}{18} \frac{1}{\N }\right)
\frac{1}{\o} +O(1)\,,\label{NDEresults}
\end{eqnarray}
and hence that (in the same notation as eq.~(\ref{photnlosingg}))
\begin{eqnarray}
\hat{\Sigma}^{gg}_{\o}&\sim & \frac{\pi \alpha_s^2 }{\V}
\left[\left(\frac{4}{15}\N -\frac{7}{18} \frac{1}{\N}\right)
\left(1-2 \frac{\asb}{\o} \ln \frac{\mu^2}{M^2}\right) +
2\frac{\asb}{\o} \left(\frac{154}{225} \N - \frac{41}{54}
\frac{1}{\N}\right)\right]+ \cdots\,,\nn\\
\hat{\Sigma}^{gq}_{\o}&\sim& \frac{\pi \alpha_s^2 }{2\N^2}
\left[\left(\frac{4}{15}\N -\frac{7}{18} \frac{1}{\N}\right)
\frac{\asb}{\o} \ln \frac{\mu^2}{M^2} + \frac{\asb}{\o}
\left(\frac{154}{225} \N - \frac{41}{54} \frac{1}{\N}\right)
\right] +\cdots \,.\label{hadnlosingq}
\end{eqnarray}

\subsection{High energy behaviour}

At high energies the total cross section for the hadroproduction
of a (heavy) quark-antiquark pair is dominated by the (Regge)
gluon fusion process of figure~\ref{hvyqrk}.
\FIGURE[b]{\epsfig{file=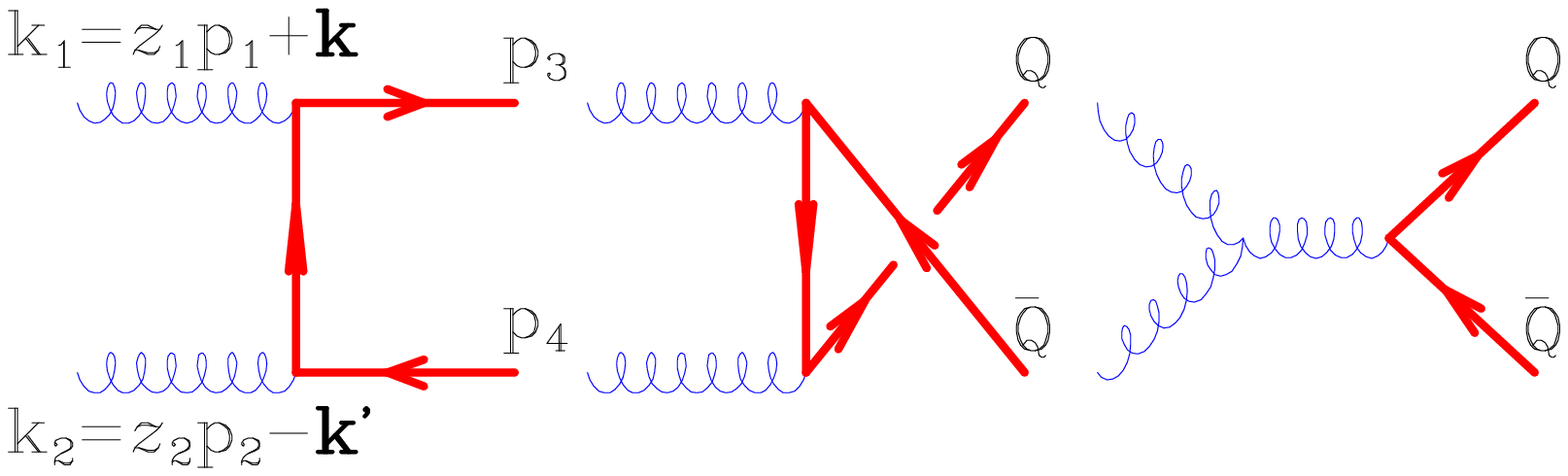,width=.6\textwidth}
\caption{Diagrams for heavy quark hadroproduction}
\label{hvyqrk}}
In the large centre of mass energy limit the perturbative
expansion of the short-distance cross section contains large
terms of order $\as^3 [\as(M^2)\log (s/M^2)]^n$ which need to be
resummed to all orders. This resummation is performed by using
the ${\k}$-factorized expression for the cross section, which
here takes the form
\begin{eqnarray}
\Sigma(\rho) &=&\sum_{ij}\int \; \f{dz_1}{z_1}\f{dz_2}{z_2}
\frac{d^2{\k}}{\pi}\frac{d^2{\k}^\prime}{\pi} \;
\hat{\Sigma}_{ij} ({\k},{\k}^\prime,M^2,z_1z_2 s)
\; \FF^i(z_1,{\k}^2,\mu^2)\times\nn\\
&&\hphantom{\sum_{ij}\int \;}\!\times\,\FF^j(z_2,{\k}^{\prime\; 2},
\mu^2)\,,\label{ktfac}
\end{eqnarray}
in which $\FF_i$ are the unintegrated parton distribution
functions and $\hat{\Sigma}$ is the (lowest-order and
gauge-invariant) off-shell continuation of the parton cross
section. Taking Mellin moments with respect to the longitudinal
momenta (see eqs.~(\ref{omegamoments}) and~(\ref{defnomegamel}))
again undoes the convolution:
\begin{equation} \label{ktfacmel}
\Sigma_\o =\sum_{ij}\int \; \frac{d^2{\k}}{\pi}\frac{d^2{\k}^\prime}
{\pi} \; H^{ij}_\o \left(\f{\k}{M},\f{{\k}^\prime}{M}\right)
\; \FF^i_\o({\k}^2,\mu^2)\FF^j_\o({\k}^{\prime\; 2},\mu^2)\,.
\end{equation}
A further simplification is obtained by Mellin transform with
respect to the transverse~momenta:
\begin{equation} \label{melfact}
\Sigma_\o =\sum_{ij} \int_C \f{d\g_1d\g_2}{(2\pi i )^2}
\left(\f{M^2}{\mu^2}\right)^{\g_1+\g_2} \FF^i_{\o}(\g_1)\;
\FF^j_{\o}(\g_2) \; H^{ij}_\o(\g_1,\g_2)\,,
\end{equation}
where the Mellin transforms of the unintegrated parton
distribution functions are defined as in eq.~(\ref{melunintpdf})
and the impact factor
\begin{equation}
H^{ij}_\o(\g_1,\g_2)=\int_0^\infty\f{d^2{\k}}{\pi {\k}^2}
\f{d^2{\k}^\prime}{\pi {\k}^{\prime 2}} \left(\f{{\k}^2}{M^2}
\right)^{\g_1}\left(\f{{\k}^{\prime 2}}{M^2}\right)^{\g_2}
H^{ij}_\o \left(\f{\k}{M},\f{{\k}^\prime}{M}\right).
\end{equation}
When we pick up the residues of the two perturbative poles
eq.~(\ref{lippole}) in the $\gamma_1$, $\gamma_2$ integrations,
we find in a sequence of steps analogous to those leading to
eq.~(\ref{gamcollfacsingglu})
\begin{equation}\label{melfacint}
\Sigma_\o = \sum_{ij} \left(\gamma_s\left(\f{\asb}{\o}\right)
\right)^2 H^{ij}_\o\left(\gamma_s\left(\f{\asb}{\o}\right),
\gamma_s\left(\f{\asb}{\o}\right)\right)
\left(\f{M^2}{\mu^2}\right)^{2\gamma_s(\asb/\o)} \;
F^{i}_{\o}(\mu^2) \; F^{j}_{\o}(\mu^2)\,,
\end{equation}
where $F^i_{\o}(\mu^2)$ are the integrated parton distribution
functions at scale $\mu^2$. \pagebreak[3]

It thus remains to determine the impact factors
$H^{ij}_\o(\gamma_1,\gamma_2)$. This requires the calculation of
the moments of the heavy quark production cross section with two
incoming off shell gluons. Details of this calculation may be
found in the appendix. The answer in our normalization, and for
$\o=0$, is
\begin{eqnarray}
H_{0}(\g_1,\g_2)&=& \frac{\pi \alpha_S^2}{\V} B(\g_1,1-\g_1)
B(\g_2,1-\g_2)\times\nn \\
&&\times \left\{4 \N \;\left[\frac{B(3-\g_1-\g_2,3-\g_1-\g_2)}
{(1-\g_1-\g_2)}\right.\right.+\nn\\
&&\hphantom{\times \Bigg\{4 \N \;\Bigg[}\!\left.+\,\frac{B(3-\g_1
-\g_2,3-\g_1-\g_2)} {(1-\g_1-\g_2)^3 \; B^2(1-\g_1,1-\g_2)}\right]-
\label{Hresult}\\
&&\hphantom{\times \Bigg\{}\!\left.-\,\frac{2}{\N} \frac{\G(2-\g_1)
\G (2-\g_2)\G (2-\g_1-\g_2)} {\G(4-2 \g_1)\G(4-2 \g_2)}
\frac{(7 -5 (\g_1+\g_2)+3 \g_1 \g_2)}{(1-\g_1-\g_2)}\right\}.\nn
\end{eqnarray}
This calculation has previously been reported
in~\cite{Camici:1996st,Camici:1997ta}. Note that
eq.~(\ref{Hresult}) is in disagreement with
ref.~\cite[eq.~(2.10)]{Camici:1996st} and
ref.~\cite[eqs.~(3.14)]{Camici:1997ta}.

When $\o=0$ $H_{\o}(\g_1,\g_2)$ has a triple pole at
$\g_1=1-\g_2$, which splits into a simple and a double pole when
$\o$ is non zero. In the vicinity of $\g_1=\g_2=1/2$ one finds
the singularity structure~\cite{Catani:1991eg}
\begin{equation} \label{mezzomezzo}
\g_1 \g_2 H_{\o}(\g_1,\g_2)= \frac{\pi \alpha_S^2}{\V} \;
\frac{\N}{6} \; \frac{1} {(1-\g_1-\g_2)}
\frac{1}{(1-\g_1-\g_2+\o)^2}\,.
\end{equation}
This singularity comes from the propagator for the $s$-channel
gluon in the third diagram in figure~\ref{hvyqrk}: it is thus
intrinsically non abelian, in the sense that it is not present in
the high energy direct photoproduction cross-section given by the
diagrams of figure~\ref{photoprod}. The singularity can drive a
substantial rise in the hadroproduction cross-section at high
energies, as we will show in detail in the next section.

Since away from these singularities $H_{\o} (\g_1,\g_2)$ is
regular in $\o$, we may write
\begin{equation}\label{Htoh}
\g_1 \g_2 H_{\o} (\g_1,\g_2)= h(\g_1,\g_2) + O(\o)\,.
\end{equation}
In the perturbative limit for small $\g_1$ and $\g_2$ we then have
\begin{equation} \label{h_nlo_exp}
h(\g_1,\g_2) \sim \f{\pi \alpha_s^2}{\V}\left[\frac{4}{15}
\N -\f{7}{18} \frac{1}{\N} +(\g_1+\g_2) \left(\f{154}{225}
\N -\f{41}{54}\frac{1}{\N} \right) + \cdots \right],
\end{equation}
so that
\begin{eqnarray}
h(0,0) &=& \f{\pi \alpha_s^2}{\V} \left(\frac{4}{15}\N -\f{7}{18}
\frac{1}{\N}\right), \label{hzz}\\
\frac{d}{d\g} h(\g,0)\Big\vert_{\g=0} &\equiv& h^\prime(0,0)=
\f{\pi \alpha_s^2}{\V} \left(\f{154}{225}\N -\f{41}{54}
\frac{1}{\N}\right).\label{hpzz}
\end{eqnarray}

As in the photoproduction case we must take care to identify the
factorization scheme correctly. In particular to change
eq.~(\ref{Hresult}) from the ``$k_T$-factorization scheme'' to
\MS\ factorization now introduces a factor of
$R(\g_1)R(\g_2)$, \pagebreak[3] with $R$ the \MS-gluon
normalization factor eq.~(\ref{Rdefinition}), so that in \MS
\begin{equation}
\g_1 \g_2 H^{gg}_{\o} (\g_1,\g_2)= h(\g_1,\g_2)R(\g_1)R(\g_2)
+ O(\o)\,.
\end{equation}
Furthermore arguments similar to those in
ref.~\cite{Catani:1994sq} may again be used to determine LLx high
energy heavy quark production cross sections with incoming quarks
in terms of those with incoming gluons: again in \MS\ (compare
eq.~(\ref{photcomsq}))
\begin{eqnarray}
\g_1 \g_2 H^{gq}_{\o} (\g_1,\g_2) &=& r_c(h(\g_1,\g_2)R(\g_1)
R(\g_2)-h(\g_1,0)R(\g_1)) + O(\o)\,,\nn\\
\g_1 \g_2 H^{qq}_{\o} (\g_1,\g_2) &=& r_c \left(h(\g_1,\g_2)R(\g_1)
R(\g_2)-h(\g_1,0)R(\g_1)\right.-\nn\\
&&\hphantom{r_c \Bigg(}\!\left. -\, h(0,\g_2)R(\g_2)+h(0,0)\right) +
O(\o)\,,
\end{eqnarray}
and similarly for the various other combinations.

Putting all this together, in the high energy limit the partonic
cross sections for hadroproduction of heavy quarks are
\begin{eqnarray}
\hat{\Sigma}^{gg}_\o&=& h\left(\g_s\left(\f{\asb}{\o}\right),
\g_s\left(\f{\asb}{\o}\right)\right)R^2\left(\g_s\left(\f{\asb}{\o}
\right)\right)\left(\frac{M^2}{\mu^2} \right)^{2 \g_s(\asb/\o)},\nn\\
\hat{\Sigma}^{g q}_\o &=& \hat{\Sigma}^{g\qbar}_\o = r_c
\left[h\left(\g_s\left(\f{\asb}{\o}\right),\g_s\left(\f{\asb}{\o}
\right)\right) R^2\left(\g_s\left(\f{\asb}{\o}\right)\right)
\left(\frac{M^2}{\mu^2} \right)^{2\g_s(\asb/\o)}\right.-\nn\\
&&\hphantom{\hat{\Sigma}^{g\qbar}_\o = r_c \Bigg[}\!\left.-\,h
\left(\g_s\left(\f{\asb}{\o} \right),0\right)R\left(\g_s
\left(\f{\asb}{\o}\right)\right) \left(\frac{M^2}{\mu^2}
\right)^{\g_s(\asb/\o)}\right],\nn\\
\hat{\Sigma}^{q \qbar}_\o&=& \hat{\Sigma}^{\qbar q}_\o =
\hat{\Sigma}^{qq}_\o = \hat{\Sigma}^{\qbar\qbar}_\o = r_c^2
\left[h\left(\g_s\left(\f{\asb}{\o}\right),\g_s\left(\f{\asb}{\o}
\right)\right)R^2\left(\g_s\left(\f{\asb}{\o}\right)\right)
\left(\frac{M^2}{\mu^2}\right)^{2\g_s(\asb/\o)}\right.-\nn\\
&&\hphantom{\hat{\Sigma}^{\qbar q}_\o = \hat{\Sigma}^{qq}_\o =
\hat{\Sigma}^{\qbar\qbar}_\o =r_c^2\Bigg[}\!-\,2h
\left(\g_s\left(\f{\asb}{\o}\right),0\right)R\left(\g_s
\left(\f{\asb}{\o}\right)\right)\left(\frac{M^2}{\mu^2}
\right)^{\g_s(\asb/\o)}+\nn\\
&&\hphantom{\hat{\Sigma}^{\qbar q}_\o = \hat{\Sigma}^{qq}_\o =
\hat{\Sigma}^{\qbar\qbar}_\o =r_c^2\Bigg[}\!\left. +\, h(0,0)
\vphantom{\left(\frac{M^2}{\mu^2} \right)^{\g_s(\asb/\o)}}\right].
\label{hxsecqq}
\end{eqnarray}

\subsection{Double leading resummation}

The perturbative expansion of the \MS\ partonic cross-sections
eqs.~(\ref{hxsecqq}) may be derived using
eqs.~(\ref{lippert}),~(\ref{Rexp}) and~(\ref{h_nlo_exp}):
\looseness=-1
\begin{eqnarray}
h\left(\g_s\left(\f{\asb}{\o}\right),\g_s\left(\f{\asb}{\o}\right)
\right)R^2\left(\g_s\left(\f{\asb}{\o}\right)\right)&\sim&
h(0,0)+2\f{\asb}{\o}h'(0,0)+O\left(\as^2\left(\f{\asb}{\o}\right)^2
\right),\nn\\
h\left(\g_s\left(\f{\asb}{\o}\right),0\right)R\left(\g_s
\left(\f{\asb}{\o}\right)\right)&\sim& h(0,0)+\f{\asb}{\o}h'(0,0)+
O\left(\as^2\left(\f{\asb}{\o}\right)^2\right),\qquad
\end{eqnarray}
with $h(0,0)$ and $h'(0,0)$ given by eqs.~(\ref{hzz})
and~(\ref{hpzz}), consistent with the fixed order results
eqs.~(\ref{hadnlosingq}). \pagebreak[3] Combining the fixed order
cross-sections with the high energy cross-sections, taking care
to subtract the terms eqs.~(\ref{hadnlosingq}) to avoid double
counting, we get
\begin{eqnarray}
\hat{\Sigma}^{g g}_\o &=& \as^2 \left\{f_{g g\; \o}^{(0)} +
4\pi\as \left[f_{g g\; \o}^{(1)}+ \ln\left(\frac{\mu^2}{M^2}\right)
\bar{f}_{g g\;\o}^{(1)} \right]\right\}\nonumber +\\
&&+\, \left[ h\left(\g_s\left(\f{\asb}{\o}\right),\g_s
\left(\f{\asb}{\o}\right)\right)R^2\left(\g_s\left(\f{\asb}{\o}
\right)\right)\left(\frac{M^2}{\mu^2} \right)^{2 \g_s(\asb/\o)}
\right.-\nonumber\\
&&\hphantom{+\, \Bigg[}\! \left. -\,h(0,0) \left(1+2 \f{\asb}{\o}
\ln \frac{M^2}{\mu^2} \right)-2 \f{\asb}{\o} h^\prime(0,0)
\right],\nn\\
\hat{\Sigma}^{g q}_\o &=& \hat{\Sigma}^{g\qbar}_\o
= \as^2 \left\{4\pi\as \left[f_{g q\; \o}^{(1)} +
\ln\left(\frac{\mu^2}{M^2}\right) \bar{f}_{g q\; \o}^{(1)}\right]
\right\}+\nonumber \\
&&\hphantom{\hat{\Sigma}^{g\qbar}_\o =}\!+\, r_c \left[h\left(\g_s
\left(\f{\asb}{\o}\right),\g_s \left(\f{\asb}{\o}\right)\right)R^2
\left(\g_s\left(\f{\asb}{\o} \right)\right)\left(\frac{M^2}{\mu^2}
\right)^{2\g_s(\asb/\o)} \right.-\nonumber \\
&&\hphantom{\hat{\Sigma}^{g\qbar}_\o =+\, r_c \Bigg[}\!-\,
h\left(\g_s\left(\f{\asb}{\o}\right),0\right)R
\left(\g_s\left(\f{\asb}{\o}\right)\right)
\left(\frac{M^2}{\mu^2} \right)^{\g_s(\asb/\o)}-\nn\\
&&\hphantom{\hat{\Sigma}^{g\qbar}_\o =+\, r_c \Bigg[}\!\left.-\,
h(0,0) \f{\asb}{\o}\ln \frac{M^2}{\mu^2}
-\f{\asb}{\o} h^\prime(0,0)\vphantom{\left(\frac{M^2}{\mu^2}
\right)^{2\g_s(\asb/\o)}}\right],\nn\\
\hat{\Sigma}^{q \qbar}_\o&=& \hat{\Sigma}^{\qbar q}_\o =
\as^2 \left\{f_{q \qbar \; \o }^{(0)} + 4\pi\as \left[
f_{q \qbar \; \o}^{(1)} +
\ln\left(\frac{\mu^2}{M^2}\right)\bar{f}_{q \qbar \; \o}^{(1)}
\right]\right\}+\nonumber \\
&&\hphantom{\hat{\Sigma}^{\qbar q}_\o = }\!+\,
r_c^2 \left[h\left(\g_s\left(\f{\asb}{\o}\right),\g_s
\left(\f{\asb}{\o}\right)\right)R^2\left(\g_s\left(\f{\asb}{\o}
\right)\right)\left(\frac{M^2}{\mu^2} \right)^{2\g_s(\asb/\o)}
\right.-\nonumber \\
&&\hphantom{\hat{\Sigma}^{\qbar q}_\o = +\,
r_c^2 \Bigg[}\!\left.-\, 2h\left(\g_s\left(\f{\asb}{\o}\right),
0\right)R\left(\g_s\left(\f{\asb}{\o}\right)\right)
\left(\frac{M^2}{\mu^2} \right)^{\g_s(\asb/\o)} + h(0,0)\right],\nn\\
\hat{\Sigma}^{qq}_\o &=& \hat{\Sigma}^{\qbar\qbar}_\o =
r_c^2 \left[h\left(\g_s\left(\f{\asb}{\o}\right),\g_s
\left(\f{\asb}{\o}\right)\right)R^2\left(\g_s\left(\f{\asb}{\o}
\right)\right)\left(\frac{M^2}{\mu^2}\right)^{2\g_s(\asb/\o)}
\right.-\nonumber\\
&&\hphantom{\hat{\Sigma}^{\qbar\qbar}_\o = r_c^2 \Bigg[}\!
\left.-\, 2h\left(\g_s\left(\f{\asb}{\o}\right),
0\right)R\left(\g_s\left(\f{\asb}{\o}\right)\right)
\left(\frac{M^2}{\mu^2} \right)^{\g_s(\asb/\o)}+ h(0,0)\right].
\end{eqnarray}
The NLLx resummation~\cite{Altarelli:2000mh} may be implemented
just as in the photoproduction case, with similar ambiguities in
the treatment of the double counting terms: here this ambiguity
should be relatively unimportant at high energies since it has no
effect on the singularities of $h(\g_s(\asb/\o),\g_s(\asb/\o))$.

\section{Asymptotic behaviour at high energy}\label{sec:4}

The aim of this paper has been to calculate and collect results
so that a matched numerical calculation of heavy quark cross
sections is possible. The matching is necessary so that the
calculation correctly includes the known fixed order results as
well as the leading tower of logarithms present at high
energy.

In the following we give some analytic results for the high energy
resummation alone; these are in no way meant to supplant the
complete analysis alluded to above, but instead serve to indicate
some features of the high energy resummation. First, we will
obtain results for the asymptotic behaviour of the gluon
distribution function at high energy, keeping the coupling fixed.
We consider two alternative choices for the anomalous dimension:
an anomalous dimension given simply by the singular term in
lowest order perturbation theory (so that $\chi \sim 1/\g$ and
has no minimum) and, for contrast, the lowest order BFKL
anomalous dimension (so that the corresponding $\chi$-function
has a minimum at $\g=1/2$). Second, we will explore the
consequences of these two forms of the anomalous dimension for the
high energy behaviour of the heavy quark photoproduction and
hadroproduction cross sections, expressing the results in terms
of the integrated gluon distribution function. For simplicity,
only the gluon contributions will be considered, since the quark
contributions are smaller. Last, we will attempt a more
quantitative estimate of the expected enhancement by letting the
coupling run and choosing $\g=\asb( \f{1}{\o}-1)$, so that
$\chi(\g)=1/(\g+\asb)$ and the momentum conservation constraint
at $\o=1$ is satisfied explicitly. This form is a good
approximation to the fully resummed anomalous
dimension~\cite{Ciafaloni:1999yw,Altarelli:2000vw,Altarelli:2000mh}
and furthermore gives a good account of scaling violations in HERA
structure function data~\cite{Ball:1994du}.

\subsection{Gluon distribution}

In $x$ space, the integrated gluon distribution at the heavy quark
mass scale, $M$, can be obtained by performing the inverse Mellin
transform of eq.~(\ref{intgluevol}) with respect to~$\o$
\begin{eqnarray}
G(x,M^2) &=& \int_C \frac{d \o}{2 \pi i} \; x^{-\o}
\left(\f{M^2}{\mu_0^2}\right)^{\gamma_s(\asb/\o)} G^0(\o)\nn \\
&=&\int_C \f{d\o}{(2 \pi i)} \exp{\left(\o L_x+\gamma_s
\left(\f{\asb}{\o}\right)l\right)} G^0(\o)\,,\label{gluondist}
\end{eqnarray}
where $L_x =\ln(1/x)$, $l=\ln(M^2/\mu_0^2)$. Here $G^0(\o)\equiv
F^g_\o(\mu_0^2)$ is a starting function at scale $\mu_0$ which we
will assume is benign in terms of poles. For simplicity, we will
work entirely in the ``$k_T$-factorization scheme'', so that the
factors of $R$ (see eq.~(\ref{Rdefinition})) are absorbed into
the gluon distribution.

The asymptotic form of $G(x,M^2)$ depends on the form of the
singular anomalous dimension $\gamma_s(\asb/\o)$. If we take the
simple form $\gamma_s(\asb/\o)=\asb/\o$, the $\o$ integration may
be done by a saddle point approximation~\cite{DeRujula:1974rf},
with the result
\begin{equation} \label{pertgluon}
G(x,M^2) \simeq \f{(\asb l)^{1/4}}{(4 \pi)^{1/2} L_x^{3/4} }
\exp \left(2 \sqrt{\asb l L_x }\right)
G^0\left(\sqrt{\f{\asb l}{L_x}}\right),
\end{equation}
where the argument of $G^0$ is the position of the saddle point.

If, on the other hand, the anomalous dimension
$\gamma_s(\asb/\o)$ is more like the BFKL form, it is convenient
to make the usual change of variables
$\o\:\:\!=\:\:\!\asb\chi(\g)$, (see eq.~(\ref{duality})),
\begin{equation}\label{gamrep}
G(x,M^2)=\int_C \f{d\g}{2\pi i}(-\asb\chi'(\g))\;
\exp{\left(\asb \chi(\g)L_x+\g l\right)}\; G^0(\asb\chi(\g))\,.
\end{equation}
If we now specialize to a $\chi$-function which has a minimum at
one half
\begin{equation}\label{chimin}
\chi(\g) \approx \chi\left(\half\right)+\half\chi^{\prime \prime}
\left(\half\right) \left(\g-\half\right)^2 + \cdots \,,
\end{equation}
so that, as $L_x$ increases, the saddle point in the $\gamma$
integration moves to $1/2$ from below. Asymptotically we then
find when $L_x\gg l$
\begin{equation} \label{BFKLgluon}
G(x,M^2)\sim \f{l\; L_x^{-3/2}}{(2\pi\asb\chi''(1/2))^{1/2}}
\exp \left(\asb L_x \chi\left(\half\right)+\half l \right)
G^0\left(\asb\chi\left(\half\right)\right).
\end{equation}

Now consider the direct component of the photoproduction
cross-section in the high energy limit. From
eq.~(\ref{gamcollfacsingglu}) we have that
\begin{equation}
\Sigma(\rho) =\int_C \f{d\o}{(2 \pi i)} \exp{\left(\o L+
\gamma_s\left(\f{\asb}{\o}\right)l\right)}
j\left(\gamma_s\left(\f{\asb}{\o}\right)\right) G^0(\o)\,,
\end{equation}
where $L\equiv\ln 1/\rho$. The function $j(\g)$, defined in
eq.~(\ref{jdef}) is free from poles in the interval $0< \g < 1$
and is therefore a smooth function in the region of the saddle
points of the integrand. It follows that at high energy for the
``perturbative'' gluon, eq.~(\ref{pertgluon}),
\begin{eqnarray}
\Sigma(\rho)&\simeq &\f{(\asb l)^{1/4}}{(4 \pi)^{1/2} L^{3/4}}\exp
\left(2 \sqrt{\asb l L}\right)\;j\left(\frac{\asb}{\os}\right)\;
G^0(\os)\nn\\
&\simeq & j\left(\frac{\asb}{\os}\right)\;G(\rho,M^2)\,,
\end{eqnarray}
where $\os = \sqrt{\asb l/L}$ is the position of the saddle
point. This expression will break down when the argument of $j$
approaches one, since $j$ has a simple pole there, but should
still be good around $\asb/\os \sim 1/2$, where it leads to a
small enhancement~\cite{Catani:1991eg}. For the BFKL gluon,
eq.~(\ref{BFKLgluon}), we find at high energy
\begin{equation}
\Sigma(\rho) \simeq j\left(\half\right) G(\rho,M^2)\,.
\end{equation}
In both cases the rise in the gluon drives the rise in the
cross-section, with only a modest enhancement over the leading
order result, $j(0) G(\rho,M^2)$.

\subsection{Hadroproduction}

We now consider the high energy behaviour of the cross section for
the hadroproduction of a pair of quarks of mass $M$. From
eq.~(\ref{melfacint}) we have
\begin{eqnarray}
\Sigma(\rho)&=& \int_C \f{d\o}{(2 \pi i)} \rho^{-\o}\;
\left(\gamma_s\left(\f{\asb}{\o}\right)\right)^2 H_\o
\left(\gamma_s\left(\f{\asb}{\o}\right),\gamma_s\left(\f{\asb}{\o}
\right)\right)\left(\f{M^2}{\mu_0^2}\right)^{2\gamma_s(\asb/\o)}
\left[G^0(\o)\right]^2 \nn \\
&=& \int_C \f{d\o}{(2 \pi i)} \exp{\left(\o L+2\gamma_s
\left(\f{\asb}{\o}\right)l\right)}\;
h_\o\left(\gamma_s\left(\f{\asb}{\o}\right),\gamma_s
\left(\f{\asb}{\o}\right)\right)\left[G^0(\o)\right]^2\nn\\
&=& \int_C \f{d\g}{2\pi i}(-\asb\chi'(\g))\;
\exp{\left(\asb \chi(\g)L+2\g l\right)} h_\o\left(\gamma,\gamma
\right) \left[G^0(\asb\chi(\g))\right]^2, \label{hadroresultg}
\end{eqnarray}
where $h_\o(\g,\g)=\g^2 H_\o(\g,\g)$, $L\equiv\ln 1/\rho$ and in
the last line we again made the change of variables
$\o=\asb\chi(\g)$. In contrast to the direct photoproduction
cross-section considered above, the asymptotic behaviour at large
$L$ will now be dominated by the singularities of
$h_{\o}(\g_1,\g_2)$ (see eq.~(\ref{mezzomezzo})),
\begin{eqnarray}
h_{\o}(\g,\g) &\simeq& \frac{A}{ (1-2\g) (1-2\g+\o)^2}\nn \\
&=&\frac{A}{\o^2}\left[\f{1}{(1-2\g)}-\f{1}{(1-2\g+\o)}
-\f{\o}{(1-2\g+\o)^2}\right]\label{beforepartfrac}
\end{eqnarray}
with $A=\pi \as^2 \frac{N_c}{6 V_c}$.

\pagebreak[3]

First consider the ``perturbative'' anomalous dimension
$\gamma=\asb/\o$. Because the behaviour in $\o$ is more intuitive
we choose to work with the $\o$ representation,
eq.~(\ref{hadroresultg}). With this choice for $\g$,
eq.~(\ref{beforepartfrac}) becomes
\begin{eqnarray}
h_{\o}\left(\f{\asb}{\o},\f{\asb}{\o}\right) &\equiv& A \;
f(\o,\o_0,\wp,\wm) \nonumber \\
&=& A\left[-\f{1}{\wp \wm (\o-\o_0)} -\f{1}{(\wp-\wm)^2}
\left\{\f{\wp}{(\o-\wp)^2}+\f{\wm}{(\o-\wm)^2} \right\}
\right.-\nonumber \\
&&\hphantom{A\Bigg[}\!\left.- \f{1}{(\wp-\wm)^3} \left\{\f{(\wm^2-3
\wm \wp)} {\wp(\o-\wp)} +\f{(3 \wm \wp-\wp^2)}{\wm (\o-\wm)}\right\}
\right],\label{splitpoles}
\end{eqnarray}
where
\begin{equation}
\o_0=2 a\,, \qquad \o_\pm=\frac{[-1\pm \sqrt{1+8 a}]}{2} \,.
\end{equation}
The dominant behaviour at large $L$ and $l$ depends on the
position of the saddle point, $\os=\sqrt{2al/L}$ relative to the
rightmost pole, $\o=\o_0$. If the saddle point is to the right of
the pole, $\os > 2 \asb$, (which means $2\asb L < l$), we
integrate along the contour $\o=\os+iz$ to obtain the result
\begin{eqnarray}
\Sigma(\rho) &\simeq & \f{A}{2 \pi}\exp{\left(\os L+2\f{\asb}{\os}
l \right)}\; \left[G^0(\os)\right]^2 \int_{-\infty}^{\infty} dz \;
f(\os+iz,\o_0,\wp,\wm)\times\nn\\
&&\hphantom{\f{A}{2 \pi}\exp{\left(\os L+2\f{\asb}{\os} l \right)}\;
\left[G^0(\os)\right]^2 \int_{-\infty}^{\infty}}\!\times\,\exp
\left(-\f{z^2}{c^2}\right),
\end{eqnarray}
where $f$ is given in eq.~(\ref{splitpoles}) and the quantity $c$,
\begin{equation}
c=\left({\f{\os ^3}{2\asb l}}\right)^{1/2}
\end{equation}
sets the scale of the gaussian fluctuations which determines
whether the poles at $\o_0,\wp$ and $\wm$ should be considered
close.

Using the saddle point method in the vicinity of poles we need to
consider integrals of the general form
\begin{equation}
K_n(y) = c^{n-1}\int_{-\infty}^{\infty} dz \f{1}{(b+iz)^n}
\exp\left(-\f{z^2}{c^2}\right),
\end{equation}
with $y=b/c$ and $b,c\neq 0$: $K_n(-y)=(-)^n K_n(y)$. An
evaluation of the first three of these integrals $K_1(y)$,
$K_2(y)$ and $K_3(y)$ may be found in the appendix~\ref{appb}. In
terms of these functions the result is
\begin{equation}
\Sigma(\rho)=\f{Ac}{(4\pi)^{1/2}}\; X(\os,\o_0,\wp,\wm)\;
\exp\big(2\sqrt{2\asb l L}\big)\; [G^0(\os)]^2\,,
\end{equation}
where
\begin{eqnarray}
X(\os,\o_0,\o_+,\o_-) &=& \f{1}{c\sqrt{\pi}} \int_{-\infty}^{\infty}
dz \; f(\os+iz,\o_0,\wp,\wm) \exp\left(-\f{z^2}{c^2}\right) \nn\\
&=&\f{1}{c \sqrt{\pi}}\left[-\f{1}{\wp\wm} K_{1}
\left(\f{(\os-\o_0)}{c}\right)\right.- \nn \\
&&\hphantom{\f{1}{c \sqrt{\pi}}\Bigg[}\!-\, K_{1}
\left(\f{(\os-\wp)}{c}\right) \f{\wm(\wm-3 \wp)}{\wp (\wp-\wm)^3}
-\nn\\
&&\hphantom{\f{1}{c \sqrt{\pi}}\Bigg[}\!-\,K_2\left(\f{(\os-\wp)}{c}
\right) \f{\wp}{c (\wp-\wm)^2}+ \nn \\
&&\hphantom{\f{1}{c \sqrt{\pi}}\Bigg[}\!+\, K_{1}
\left(\f{(\os-\wm)}{c}\right) \f{\wp(\wp-3\wm)}{\wm (\wp-\wm)^3}
-\nn\\
&&\hphantom{\f{1}{c \sqrt{\pi}}\Bigg[}\!\left.-\,K_2
\left(\f{(\os-\wm)}{c}\right) \f{\wm}{c (\wp-\wm)^2} \right].
\label{Xdef}
\end{eqnarray}
Expressed in terms of the gluon distribution
eq.~(\ref{pertgluon}), we find
\begin{equation}\label{pertxsect}
\Sigma(\rho) \simeq A\f{\pi^{1/2}L^{3/4}}{(2\asb l)^{1/4}}\;
X(\os,\o_0,\wp,\wm)\; \left[G(\sqrt{\rho},M^2)\right]^2,
\end{equation}
where we have used the fact that if $x=\sqrt{\rho}$, $L_x=\half L$
and thus $2\sqrt{2\asb l L}=4\sqrt{2\asb l L_x}$. As
$\os\rightarrow\o_0$ from above, the enhancement factor
$X(\os,\o_0,\wp,\wm)$ grows until it reaches a maximum value at
the pole.

When $\os$ passes the pole, we have to add the residue of the pole
to the saddle contribution. The result for the pole contribution
for $2\asb L > l$ is
\begin{eqnarray}
\Sigma(\rho) &\simeq& \f{A}{2 a} \exp\left(2\asb L +l\right)\;
[G^0(\o_0)]^2 \nn\\
&\simeq& A \f{\pi l}{4 \asb^3} \exp\left(2\asb L -l\right)\;
[G(e^{-l/4\asb},M^2)]^2\,.\label{poledominated}
\end{eqnarray}
At high energies this contribution is dominant, since it grows as
$e^{2aL}=\rho^{-2\asb}$. Taken at face value,
eq.~(\ref{poledominated}) thus indicates a dramatic growth of the
hadroproduction cross-section with increasing energy, driven not
by the growth of the gluon distributions but rather by the
perturbative singularity in the partonic cross-section.

If instead we consider the BFKL anomalous dimension, the
asymptotic behaviour of the hadroproduction cross-section is best
determined using the $\g$-representation, eq.~(\ref{hadroresultg})
and $h_\o$ as given by eq.~(\ref{beforepartfrac}), The function
$\chi(\g)$, eq.~(\ref{chimin}), has a minimum at $\g=1/2$, and
$\Sigma$ is thus approximately given by
\begin{equation}
\simeq \f{\asb A \chi''(1/2)}{8} \int_C
\f{d\g}{2\pi i} \; \f{\exp{\left(\asb \chi(\g)L+2\g l\right)}}
{(\g-\half-\half a\chi(1/2))^2}\; [G^0(\asb\chi(\g))]^2\,.
\end{equation}
The integrand now has a saddle point at
\begin{equation}
\gs = \half - \f{2 l}{\asb L \chi^{\prime \prime}(1/2)}\,,
\end{equation}
and the fluctuations about the saddle are determined by
$c_\gamma=\sqrt{2/(a \chi^{\prime \prime} L)}$. The saddle point
is always to the left of $1/2$ which it approaches
asymptotically as $L$ increases. The double pole is at a fixed
distance to the right of $1/2$. The behaviour at large
$L$ is thus given by
\begin{eqnarray}
\Sigma(\rho)&\simeq& \f{a A \sqrt{\chi''(1/2)} }{16 \pi c_\gamma}
K_2\left(\f{a \chi(1/2)}{2 c_\gamma}\right)
\exp{\left(\asb\chi\left(\half\right)L+l\right)}
\left[G^0\left(\asb\chi\left(\half\right)\right)\right]^2\nn \\
&\simeq&\f{A \sqrt{\chi''(1/2)}}{2 \asb [\chi(1/2)]^2
\sqrt{2 \pi a L}} \exp{\left(\asb\chi\left(\half\right)L+l
\right)}\; \left[G^0\left(\asb\chi\left(\half\right)\right)
\right]^2
\end{eqnarray}
which when expressed in terms of the BFKL gluon
eq.~(\ref{BFKLgluon}) becomes
\begin{equation} \label{BFKLxsectcat}
\Sigma(\rho) \simeq \f{A}{16} \sqrt{\frac{2\pi}{\asb\chi''(1/2)}}
\left(\f{\chi''(1/2)}{\chi(1/2)}\right)^2
\f{L^{5/2}}{l^2}\left[G(\sqrt{\rho},M^2)\right]^2.
\end{equation}
Clearly in the BFKL case, although there is some
enhancement~\cite{Catani:1991eg}, it is rather less dramatic than
in the previous case. This is because when $\chi(\g)$ has a
minimum, the saddle point cannot travel past the pole at
$\g=1/2$.

In summary, because of the singularity structure of
eq.~(\ref{mezzomezzo}) substantial enhancements of the
hadroproduction cross-section are possible at high energy, even
when the gluon distribution function does not have the singular
behaviour predicted by lowest order BFKL.

\subsection{A more realistic model}

In order to attempt a quantitative estimate of the enhancement in
the hadroproduction cross-section at the Tevatron and LHC, we
take a simple model for the anomalous dimension function which is
close to \pagebreak[3] the one-loop value and satisfies
momentum~conservation,
\begin{equation} \label{oneloopg}
\g(\o) = \asb \left(\f{1}{\o}-1\right).
\end{equation}
We also include running coupling effects. In a theory with a
running coupling the product $l \asb $ is replaced by
\begin{equation} \label{xidef}
\zeta(\mu)= \f{\N}{\pi }\int_{\mu_0^2}^{\mu^2} \f{d k^2}{k^2}
\alpha_S(k^2) = \f{\N}{\pi b_0} \ln \left(\f{\ln(\mu^2/\Lambda^2)}
{\ln(\mu_0^2/\Lambda^2)} \right).
\end{equation}
If we take the leading order parameter $\Lambda=200\,\mbox{MeV}$
(corresponding to the leading order coupling at the $Z$-mass
$\as(M_z)=0.123$), we find that $a=0.224$. Further, taking
$\mu_0=1\,\mbox{GeV}$ from eq.~(\ref{xidef}) we find
$\zeta(M)\simeq 1$. We shall adopt this value in our numerical
estimates below, although the details will change if we use
different values for $\mu_0$ and $\Lambda$.

The modified anomalous dimension, eq.~(\ref{oneloopg}), only
changes eq.~(\ref{pertgluon}) in a minor~way
\begin{equation}\label{realglue}
G(x,\mu^2) = \f{(\zeta(\mu))^{1/4}}{(4 \pi)^{1/2} L_x^{3/4} }
\exp \left[2 \sqrt{\zeta(\mu) L_x } - \zeta(\mu)\right]
\; G^0\left(\sqrt{\f{\zeta(\mu)}{L_x}}\right).
\end{equation}
More interesting is the effect on the hadroproduction
cross-section. $h_\o$ still has the form given by
eq.~(\ref{splitpoles}) but the positions of the single and double
poles are now given by
\begin{eqnarray}
\o_0 &=&\frac{2 \asb}{(1+2\asb)}\,, \nn\\
\o_\pm &=& -\f{1}{2}-\asb \pm \sqrt{\left(\f{1}{2}+\asb\right)^2 +
2 \asb}\,.
\end{eqnarray}
The saddle point is now at
\begin{equation}
\os = \left({\f{2 \zeta(M)}{L}}\right)^{1/2}.\label{realisticsp}
\end{equation}
Some plausible numerical values of the parameters are given in
table~\ref{Saddletable}. The leading pole is at $\o=\o_0$,
closely followed by the double pole $\o_{+}$, while $\o_-$ is in
practice of little consequence.
\TABLE[t]{\centerline{
\begin{tabular}{|c|c|c|}\hline
& Tevatron & LHC \\ \hline
$L=\ln ({S}/{4 M^2})$ & 10.6 & 14.5 \\
$\zeta(M) $ & 1.0 & 1.0 \\
$\os= \big(2 \zeta(M)/L\big)^{1/2}$ & 0.43 & 0.37 \\
\hline
$\o_0 =2\asb/(1+2 \asb)$ & 0.31 & 0.31 \\
$\wp= -\f{1}{2}-\asb + \big((\f{1}{2}+\asb)^2+2\asb \big)^{1/2}$
& 0.26 & 0.26 \\
$\wm= -\f{1}{2}-\asb - \big((\f{1}{2}+\asb)^2+2\asb \big)^{1/2}$
& $-1.71$ & $-1.71$ \\
\hline
$c = [\os^3/2\zeta(M)]^{1/2}$ & 0.20 & 0.16 \\
\hline
$|(\o_0-\os)/c|$ & 0.61 & 0.39 \\
$|(\o_+-\os)/c|$ & 0.85 & 0.68 \\
$|(\o_--\os)/c|$ & 10.6 & 13.0 \\
\hline
$X(\os,\o_0,\o_+,\o_-)$ & 1.26 & 2.42 \\
\hline
\end{tabular}
}
\caption{Values of parameters at Tevatron and LHC, taking
$M=5\,\mbox{GeV}$, $\asb(M^2)=0.224$}\label{Saddletable}}

We see from the table that $\os >\o_0$ at both the Tevatron and
LHC, so we will concentrate on this case. The more dramatic
behaviour, eq.~(\ref{poledominated}) only seems to set in at yet
higher energies, of the order of a few hundred TeV. The scale of
the gaussian fluctuations which determines the closeness of the
poles is now set by
\begin{equation}
c=\left({\f{\os^3}{2 \zeta(M)}}\right)^{1/2}.
\end{equation}
The saddle point result is (using eqs.~(\ref{splitpoles})
and~(\ref{Xdef}))
\begin{eqnarray}
\Sigma(\rho) &\simeq & A\left(\f{\os^3}{8\pi\zeta(M)}\right)^{1/2}
\exp\left(2 \sqrt{2 \zeta(M) L}-2\zeta(M)\right)
X(\os,\o_0,\o_+,\o_-)\times\nn\\
&&\times\, [G^0(\os)]^2\,.\label{ANS}
\end{eqnarray}
We can express eq.~(\ref{ANS}) in terms of the gluon distribution
function, eq.~(\ref{realglue}) and~obtain
\begin{equation} \label{pertxsectr}
\Sigma(\rho) \simeq A \f{\pi^{1/2}L^{3/4}}{(2\zeta(M))^{1/4}}
X(\os,\o_0,\o_+,\o_-)\left[G(\sqrt{\rho},M^2) \right]^2.
\end{equation}

In order to estimate a $K$-factor we calculate the LO cross
section with
\begin{equation}
\g_1 \g_2 H_\o(\g_1,\g_2)\longrightarrow h(0,0)
\end{equation}
where (cf.\ eq.~(\ref{hadnlosingq})) $h(0,0)$ is given by
eq.~(\ref{hzz}). The result is
\begin{equation}
\Sigma(\rho) = h(0,0) \f{\pi^{1/2}L^{3/4}}{(2 \zeta(M))^{1/4}}
\big[ G(\sqrt{\rho},M^2) \big]^2,
\end{equation}
and hence the $K$-factor is
\begin{equation}
K = 1+\f{A}{h(0,0)}
\big[X-1\big],
\end{equation}
where $A/h(0,0) = 135/181\simeq 0.75$. With the parameters in
table~\ref{Saddletable}, the numerical value of $K$ is thus
around 1.4 (2) at the Tevatron (LHC). These numbers indicate that
resummation \emph{can} give rise to sizeable enhancements. A full
numerical analysis is required to show whether they actually do.

\section{Conclusion}\label{sec:5}

We have presented a matched formalism which may be used to provide
accurate estimates of the heavy quark cross section at high
energy. Simple analytic calculations suggest that the high energy
resummation can give large effects, even with a gluon
distribution which has the form dictated by lowest order
perturbation theory. The cause of this enhancement is an
$s$-channel gluon which leads to the singular behaviour of the
non-abelian contribution to the impact factor in
eq.~(\ref{mezzomezzo}). The effect is therefore universal, in the
sense that at high energies it will provide the leading
contribution to hadroproduction, resolved photoproduction and
even double resolved $\g\g$ heavy quark production. The
persistence of the singular behaviour in yet higher orders
remains an open question. Furthermore, detailed numerical work
will be required to establish the quantitative significance of
the enhancement at existing and future colliders.

\acknowledgments

One of us (RKE) would like to thank W.A.~Bardeen for encouragement
and advice. This work was supported in part by the U.S. Department
of Energy under Contract No. DE-AC02-76CH03000, and by EU TMR
contract FMRX-CT98-0194 \mbox{(DG 12-MIHT).}

\paragraph{Note added in proof.}

After the appearance of this paper, we were informed by Marcello
Ciafaloni that he is aware that some of the equations in
refs.~\cite{Camici:1996st,Camici:1997ta} are indeed incorrect. An
erratum correcting some of the equations in ref.~\cite{Camici:1997ta}
has now been submitted to Nuclear Physics B in May 2001.  The result
of Camici and Ciafaloni for the hadronic impact factor as corrected by
their erratum to ref.~\cite{Camici:1997ta} is now in agreement with
our eq.~(\ref{Hresult}).

We would like to thank Marcello Ciafaloni for correspondence 
on this matter.

\appendix

\section{The hadronic impact factor}\label{appa}

We wish to calculate the off-shell cross section (impact factor)
for the production of a heavy quark pair by a gluon fusion process,
\begin{equation} \label{QQprod}
g(\ku)+ g(\kd) \rightarrow Q(\pt) + \bar{Q}(\pq)\, .
\end{equation}
The incoming and outgoing momenta are specified with respect to
lightlike vectors $p_1$ and $p_2$.  We perform the Sudakov expansion
of all four momenta, where ${\k},{\k}^\prime$ and $\dt$ are euclidean
vectors in the plane transverse to $\pu$ and $\pd$,
\begin{eqnarray} 
\label{sudakov}
k_1& = & z_1 p_1 + {\k} \, ,  
\nn \\ 
k_2& = & z_2 p_2 - {\k}^\prime \, ,  
\nn \\
p_3& = & (1-x_1) z_1 p_1 + x_2 z_2 p_2 + {\k} - \dt \, , 
\nn \\
p_4& = & x_1 z_1 p_1 + (1-x_2) z_2 p_2 - {\k}^\prime  + \dt \, .
\end{eqnarray}
The specific form of the incoming momenta, $k_1$ and $k_2$ in
eq.~(\ref{sudakov}) gives the dominant contribution at high energy and
leads to a gauge invariant cross
section~\cite{Collins:1991ty,Catani:1991eg}.  The matrix element
squared, $|{\cal A}|^2$ is calculated taking the polarization sums of
the incoming gluons to be,
\begin{equation}
\label{polvecs}
\sum_\lambda \varepsilon^{\mu}_\lambda (k_1) \varepsilon^{*
\nu}_\lambda(k_1)= \f{{\k}^\mu {\k}^\nu}{{\k}^2}\,,
\qquad 
\sum_\lambda
\varepsilon^{\mu}_\lambda (k_2) \varepsilon^{* \nu}_\lambda(k_2)=
\f{{\k}^{\prime \mu} {\k}^{\prime \nu}}{{\k}^{\prime 2}}\,.
\end{equation}
 
The dimensionless partonic hard cross section $\hat{\Sigma}$ is
obtained from the matrix element, ${\cal A}$, by integrating over the
phase space, including the flux factor, ($2 \nu \equiv 4 z_1 z_2 p_1
\cdot p_2$), and multiplying by the square of the heavy quark mass,
$M^2$,
\begin{equation} 
\label{Goldenrule}
\hat{\Sigma}({\k},{{\k}^\prime},M^2,\nu)= \frac{M^2}{2\nu} \int \,
d\Phi^{(2)} \, |{\cal A}({\ku},{\kd},p_3,p_4)|^2 \, .
\end{equation}
Our aim in this appendix is to calculate the arbitrary moments of this
cross-section with respect to the transverse momenta ${\k^{2}}$ and
${\k^{\prime 2}}$ and the zeroth moment with respect to the variable
$\nu$. We define the moments of the cross section as
\begin{eqnarray} 
H_\o(\g_1,\g_2) &=& \int_0^\infty \frac{d^2{\k}}{\pi {\k}^2}
\int_0^\infty \frac{d^2 {\k}^\prime}{\pi {\k}^{\prime \, 2}}
\left(\frac{{\k}^2}{M^2} \right)^{\g_1}
\left(\frac{\k^{\prime \, 2}}{M^2} \right)^{\g_2} \times
\nn \\ &&
	\hphantom{\int_0^\infty}\!
\times\, \int_0^\infty \frac{d\nu}{\nu} \left(\frac{4
M^2}{\nu}\right)^\o \hat{\Sigma} ({\k},{\k}^\prime,M^2,\nu) \, .
\label{Fundamentalquantity}
\end{eqnarray}
The results of this calculation for $\o=0$ have been been reported in
refs.~\cite{Camici:1996st,Camici:1997ta}.  We are in disagreement with
the final results reported in these references.\footnote{Explicitly,
we disagree with eq.~(2.10) of ref.~\cite{Camici:1996st} and
eqs.~(3.14a), (3.14b) of ref.~\cite{Camici:1997ta}.  In addition, we
believe that the description of the calculation in
ref.~\cite{Camici:1997ta} contains errors or typographical mistakes in
the final results cited above, in eqs.~(3.3), (3.7), (3.10a), (3.10b)
and in the appendix eqs.~(A.1b), (A.1c), (A.5), (A.6), (A.9), (A.10)
and (A.16).}  We follow the general calculational procedure of
ref.~\cite{Camici:1997ta} and present this appendix because we believe
that a complete description of the correct calculation is a useful
addition to the literature.

We start with $|{\cal A}|^2$, the (off-shell) squared matrix element
for the gluon fusion $Q\bar{Q}$ production process given in
eq.~(\ref{QQprod}) calculated with polarization vectors,
eq.~(\ref{polvecs}). In an explicitly Lorentz invariant notation we
have,
\begin{eqnarray} 
\label{Catani}
|{\cal A} (\ku,\kd,p_3,p_4)|^2 &=&  g^4 
\Biggl[ \frac{1}{2 \N} {\cal A}^{(ab)} +\frac{\N}{2 \V} {\cal A}^{(nab)}\Biggr],
\\
{\cal A}^{(ab)}&=&4 \Biggl[\frac{(\pd \ku) (\pu \kd)}{(\pu \pd)^2}\Biggr]^2 
\Biggl\{ \frac{(\pu \pd)^2}{(t-M^2) (u-M^2)}-
\nonumber \\ && 
-\,\frac{1}{\ku^2 \kd^2} 
\Biggl[(\pu \pd)+2 \frac{(\pu \pq) (\pd \pt)}{(t-M^2)}
+2 \frac{(\pu \pt) (\pd \pq)}{(u-M^2)}\Biggr]^2 \Biggr\}\,,
\\
{\cal A}^{(nab)}&=&\Biggl[ \frac{(\pd \ku) (\pu \kd)}{(\pu \pd)^2} \Biggr]^2
\Biggl\{ 4 (\pu \pd)^2 \Biggl[ -\frac{1}{ (t-M^2) (u-M^2)}-
\nonumber \\ &&
	\qquad\qquad
-\,\frac{1}{s} \Biggl( \frac{1}{ (t-M^2)}-\frac{1}{(u-M^2)}\Biggr) 
\Biggl(\frac{(\pu \pt)}{(\pu \kd)}-\frac{(\pd \pt)}{(\pd \ku)}\Biggr)+
\nonumber\\&&
	\qquad\qquad
+\,\frac{1}{s} \frac{(\pu \pd)}{(\pu \kd) (\pd \ku)} \Biggr] +
\nonumber \\ && 
	\qquad
+\,\frac{2}{ \ku^2 \kd^2}
\Biggl[ (\pu \pd)+4 \frac{(\pu \pq) (\pd \pt)}{(t-M^2)}
-\frac{\Delta}{s}\Biggr]\times
\nonumber \\&&
\qquad\qquad
\times\Biggl[(\pu \pd)
 +4 \frac{(\pu \pt) (\pd \pq)}{(u-M^2)}+\frac{\Delta}{s} \Biggr]\Biggr\},
\end{eqnarray}
where 
\begin{equation} 
\label{Mandel}
s=(k_1+k_2)^2\,,\qquad t=(k_1-p_3)^2\,, \qquad u=(k_1-p_4)^2\, , 
\end{equation}
and we have defined
\begin{eqnarray} 
\label{Catani_last}
\Delta&=&2 (\pu \pd) \Biggl[ 2 \frac{(\pu \pt) (\pd \pq)}{(\pu \pd)}
-2 \frac{(\pu \pq) (\pd \pt)}{(\pu \pd)} -\ku^2 \frac{(\pu \pq)}{(\pu
\kd)} +
\nonumber \\&&
	\hphantom{ 2(\pu \pd)\Biggl[}\!
+ \kd^2 \frac{(\pd \pq)}{(\pd \ku)} +(\pq \ku)-(\pq \kd)\Biggr]\, .
\end{eqnarray}
eqs.~(\ref{Catani})--(\ref{Catani_last}) are identical to the expression
of Catani, Ciafaloni and Hautmann~\cite{Catani:1991eg} except that we
have performed the interchange $p_3 \leftrightarrow p_4$, necessary in
our notation, eq.~(\ref{QQprod}). It is also equivalent to the
expression of Collins and Ellis~\cite{Collins:1991ty}.

Using the Sudakov parametrization of the final state momenta,
eq.~(\ref{sudakov}), we can write the two-body phase space of the
produced $Q\bar{Q}$ pair as follows,
\begin{eqnarray} 
d\Phi^{(2)} & =& \f{\nu}{8\pi^2}\, dx_1 \, dx_2 \, d^2\dt \,
\delta(x_2(1-x_1)\nu-({\k}-\dt)^2-M^2) \times\qquad
\nn\\&&
\times \delta(x_1(1-x_2)\nu-(-{\k}^\prime+\dt)^2-M^2) 
\label{parm0}\\
&\equiv & \f{1}{8\pi^2}\f{dx_1}{x_1(1-x_1)} \, d^2\tilde{\dt} \,
\delta\Big(\nu-\f{\tilde{\dt}^2+M^2}{x_1(1-x_1)}-\q^2\Big) 
\label{parm1}\\
& \equiv & \f{1}{8\pi^2}\f{dx_2}{x_2(1-x_2)} \, d^2\hat{\dt} \,
\delta\Big(\nu-\f{\hat{\dt}^2+M^2}{x_2(1-x_2)}-\q^2\Big),
\label{parm2}
\end{eqnarray}
where
\begin{equation}
\begin{array}[b]{rclcrcl}
\q&=&{\k}-{\k}^\prime\,, &\qquad&
\nu&=&z_1z_2(2p_1\cdot p_2)=s+\q^2\,, 
\\[3pt]
\tilde{\dt}&=&\dt-{\k} x_1-{\k}^\prime(1-x_1)\,,
&\qquad& 
\hat{\dt}&=&\dt-{\k}^\prime x_2-{\k}(1-x_2)\,.
\end{array}
\end{equation}

It proves convenient to use both parametrizations of the phase-space
as given in eqs.~(\ref{parm1}) and (\ref{parm2}). Different terms in
the integrand are more easily integrated using one or the other forms
of the phase space.  For the phase space parameterized as in
eq.~(\ref{parm1}) we find that
\begin{eqnarray}
s &=& \f{\tilde{\dt}^2+M^2}{x_1(1-x_1)},
\nn\\
M^2-t &=& x_1(s+{\k}^2)+(1-x_1){\k}^{\prime \, 2}
 +2\tilde{\dt}\cdot{\k}^\prime,
\nn\\
M^2-u &=& (1-x_1)(s+{\k}^2)+x_1{\k}^{\prime \, 2}
 -2\tilde{\dt}\cdot{\k}^\prime,
\nn\\
x_2 &=& \f{\big[(1-x_1)\q-\tilde{\dt}\big]^2+M^2}{(1-x_1)\nu},
\nn\\
1-x_2 &=& \f{\big[x_1 \q+\tilde{\dt}\big]^2+M^2}{x_1 \nu},
\end{eqnarray}
whereas for the phase space as described by eq.~(\ref{parm2}) we have
\begin{eqnarray}
s &=& \f{\hat{\dt}^2+M^2}{x_2(1-x_2)},
\nn\\
M^2-t &=& x_2(s+{\k}^{\prime \, 2})+(1-x_2){\k}^2+2\hat{\dt}\cdot{\k},
\nn\\
M^2-u &=& (1-x_2)(s+{\k}^{\prime \, 2})+x_2{\k}^2-2\hat{\dt}\cdot{\k},
\nn\\
x_1 &=& \f{\big[(1-x_2)\q+\hat{\dt}\big]^2+M^2}{(1-x_2) \nu},
\nn\\
1-x_1 &=& \f{\big[x_2\q-\hat{\dt}\big]^2+M^2}{x_2 \nu} \, .
\end{eqnarray}

To simplify the formulae we define a reduced matrix element squared,
$D$
\begin{equation} 
\label{DDef}
|{\cal A}(\ku,\kd,p_3,p_4)|^2 = 
 \frac{g^4 \nu^2}{2 \V} D({\k},{\k}^\prime,\dt,x_1,M^2)\, .
\end{equation}
Expressed in terms of the reduced matrix element, $D$, and after the
inclusion of the other factors from eq.~(\ref{Goldenrule}), the Mellin
transform of impact factor, eq.~(\ref{Fundamentalquantity}), takes the
form
\begin{eqnarray} 
\label{Hequation}
H_\o(\g_1,\g_2)&=& \frac{\alpha_S^2 M^2}{2 \V} \int\, d\nu
\Big(\frac{4 M^2}{\nu} \Big)^{\o} \, \frac{d^2{\k}}{\pi {\k}^2} \,
\frac{d^2{\k}^\prime}{\pi {\k}^{\prime 2}} \, dx_1 \, dx_2\,
d^2{\dt}\times
\nn \\&&
\times  \nu  \delta(x_2(1\!-\!x_1)\nu-({\k}\!-\!\dt)^2\!-M^2)
\delta(x_1(1\!-\!x_2)\nu-(-{\k}^\prime\!+\dt)^2\!-M^2)\! \times
\nn\\&&
\times  \left(\frac{{\k}^2}{M^2}\right)^{\g_1}
\left(\frac{\k^{\prime 2}}{M^2}\right)^{\g_2}
D({\k},{\k}^\prime,\dt,x_1,M^2) \, .
\end{eqnarray}

We now perform a further separation of the reduced matrix element,
eq.~(\ref{DDef}), to isolate the sub-leading and leading colour
pieces, corresponding to the abelian and non-abelian contributions
respectively,
\begin{eqnarray}
D&=& (D^{(1)}+D^{(2)}+D^{(3)})\,,
\\
D^{(1)}&=&\frac{1}{\N}\left[\frac{-1}{(M^2-u)(M^2- t)}+
\frac{(B+C)^2}{{\k^2}\k^{\prime 2}}\right],  
\nn \\
D^{(2)}&=&\N\left[\frac{1}{s}\left(\frac{1}{M^2-t}-\frac{1}{M^2-u}\right)
(1-x_1-x_2)-\frac{B^2+C^2}{{\k^2}\k^{\prime 2}}+\right.
\nn\\&&
	\hphantom{\N\Biggl[}\!
\left. +\frac{2(B-C)}{{\k^2} {\k^{\prime 2}} s}\left(
(1-x_2) {\k^2} 
+(1-x_1) {\k^{\prime 2}} +{\k} \cdot{\k^\prime}\right)\right],
\nn \\
D^{(3)}&=&\N\left[\frac{2}{\nu s}-\frac{2}{{\k^2}\k^{\prime 2}}
\frac{\left((1-x_2) {\k^2}+(1-x_1){\k^{\prime 2}}
+{\k}\cdot{\k}^\prime\right)^2}{s^2}\right] .
\label{Deqn}
\end{eqnarray}
The quantities $B$ and $C$ are given by
\begin{eqnarray}
B&=&\frac{1}{2}-\frac{(1-x_1)(1-x_2)\nu}{M^2-u}+\frac{\nu(1-x_1-x_2)}{2s}+
\frac{\dt\cdot({\k}+{\k}^\prime)}{s}\,,
\\
C&=&\frac{1}{2}-\frac{x_1x_2\nu}{M^2- t}-\frac{\nu(1-x_1-x_2)}{2s}-
\frac{\dt\cdot({\k}+{\k}^\prime)}{s}\,, 
\end{eqnarray}
and the Mandelstam variables of the hard sub-process are given by
eq.~(\ref{Mandel}). The non-abelian part has been further split into
two terms, $D^{(2)}$ and $D^{(3)}$, each of which leads to a finite
$\dt$ integration.  The integration over ${\dt}$ is finite because of
the gauge cancellations between the different terms in $B$ and $C$
which ensures that $B,C \rightarrow {\dt}^{-2}$ at large ${\dt}$.
Note that eqs.~(\ref{Deqn}) are finite as $\k^2$ and $\k^{\prime 2}$
tend to zero and that the singularities at small $\k^2$ and
$\k^{\prime 2}$ are only apparent. This can be established by putting
the expression over a common denominator and using the identity (cf.\
eq.~(\ref{parm0}))
\begin{equation}
x_2 \nu = x_1 \nu+{\k}^2-{\k}^{\prime 2}-2 \, {\k} \cdot {\dt} 
+2 \, {\k^{\prime}} \cdot {\dt}\,.
\end{equation}

The first step in the evaluation of the $\o=0$ limit of
eq.~(\ref{Hequation}) is to use the $\delta$-function in
eq.~(\ref{parm1}) or~(\ref{parm2}) to eliminate the variable
$\nu$. The choice of which parametrization of the phase space to use
is made so that the subsequent integration over $\dt$ contains at most
one non-trivial azimuthal integration~\cite{Camici:1997ta}.

\pagebreak[3]

After combining the denominators using Feynman parameters and shifting
the variable of integration in the normal way, the integration over
$\dt$ can be performed.  We next perform the ${\k}_1^2$ and ${\k}_2^2$
integrations, which can all be evaluated using the following
identities,
\begin{eqnarray} 
&&\int_0^\infty\f{d^2{\k}}{\pi{{\k}^2}}\left(\f{{{\k}^2}}{M^2}\right)^\g
\f{1}{(M^2+\xi{{\k}^2})^n}=(M^2)^{-n}\xi^{-\g}B(\g,n-\g)\, ,
\label{integral1}\\
&&\int_0^\infty\f{d^2{\k}_1}{\pi {\k}_1^2}\f{d^2{\k}_2}{\pi{\k}_2^2}
\left(\f{{\k}_1^2}{M^2}\right)^{\g_1}
\left(\f{{\k}_2^2}{M^2}\right)^{\g_2}
\f{1}{(M^2+\xi_1{{\k}_1^2}+\xi_2{{\k}_2^2})^n}=
\nn\\
&&\qquad\qquad\qquad\qquad= (M^2)^{-n}\xi_1^{-\g_1}\xi_2^{-\g_2}
\f{\G(\g_1)\G(\g_2)\G(n-\g_1-\g_2)}{\G(n)}\, ,
\label{integral2}\\
&&\int_0^\infty\f{d^2{\k}_1}{\pi {\k}_1^2}\f{d^2{\k}_2}{\pi{\k}_2^2}
\left(\f{{\k}_1^2}{M^2}\right)^{\g_1}
\left(\f{{\k}_2^2}{M^2}\right)^{\g_2}
\f{1}{[M^2+\xi({{\k}_1+{\k}_2)^2}]^n}=
\nn\\
&&\qquad\qquad= (M^2)^{-n}\xi^{-\g_1-\g_2}
\f{\G(n-\g_1-\g_2)\G(1-\g_1-\g_2)\G(\g_1)\G(\g_2)}{\G(n)\G(1-\g_1)\G(1-\g_2)}\,.\qquad
\label{integral3}
\end{eqnarray}
Finally we perform the integration over the Feynman parameter and over
the variable $x_1$ or $x_2$.
                           
In the $\o=0$ limit, the result for the abelian piece is
\begin{eqnarray}
H^{(1)}(\g_1,\g_2)&=& -\frac{\pi \alpha_s^2}{\V} \frac{2}{\N} 
\G(1-\g_1-\g_2)\G(\g_1)\G(\g_2) \times
\nn \\&&
\times \left[\f{B(1-\g_1,1-\g_1)B(1-\g_2,1-\g_2)}{4}-\right.
\nn\\&&
	\hphantom{\times\Biggl[}\!
\left.-\f{B(1-\g_1,2-\g_1)B(1-\g_2,2-\g_2)}{(3-2\g_1)(3-2\g_2)}
\left(1+(1-\g_1)(1-\g_2)\right)\right] 
\nn \\
& \equiv &-\frac{\pi \alpha_s^2}{\V} \frac{2}{\N}
\G(1-\g_1-\g_2)\G(\g_1)\G(\g_2) \frac{\pi}{64} \frac{\G(1-\g_1)
\G(1-\g_2)} {\G(\frac{5}{2}-\g_1) \G(\frac{5}{2}-\g_2)}\times
\nn \\ &&
\times\, 4^{\g_1+\g_2} (7-5 (\g_1+\g_2)+3 \g_1 \g_2)
\nn \\
& \equiv &
-\frac{\pi \alpha_s^2}{\V} \frac{2}{\N} B(\g_1,1-\g_1)B(\g_2,1-\g_2)
\G(2-\g_1-\g_2) \times
\nn \\&&
\times \frac{\G(2-\g_1) \G (2-\g_2)}
{ \G(4-2 \g_1)\G(4-2 \g_2)} 
\frac{(7-5 (\g_1+\g_2)+3 \g_1 \g_2)}{(1-\g_1-\g_2)}\, , 
\end{eqnarray}
where we have written the same expression in three ways in order to
facilitate comparison with the wrong expression of
ref.~\cite{Camici:1997ta} and the correct expression of
ref.~\cite{Catani:1991eg}. The last form is the form quoted in this
paper.  In the $\o=0$ limit, the result for the non-abelian pieces is,
\begin{eqnarray}
H^{(2)}(\g_1,\g_2)&=&\frac{\pi \alpha_s^2}{\V}
4 \N \frac{B(3-\g_1\!-\g_2,3-\g_1\!-\g_2)} {(1-\g_1-\g_2)}
B(\g_1,1-\g_1)B(\g_2,1-\g_2)\,,\qquad~~
\\ 
\label{Hthree}
H^{(3)}(\g_1,\g_2)&=& \frac{\pi \alpha_s^2}{\V}
4 \N \frac{B(3-\g_1\!-\g_2,3-\g_1\!-\g_2)} {(1-\g_1-\g_2)^3}
\frac{\G(\g_1) \G(\g_2) \Gamma^2(2-\g_1\!-\g_2)}
{\Gamma(1 -\g_1) \Gamma(1 -\g_2)}\,. \qquad~~
\end{eqnarray}
The final result for $H$ is given by the sum of all three contributions
\begin{equation} 
\label{Htot}
H_{0}(\g_1,\g_2)=H^{(1)}(\g_1,\g_2)+H^{(2)}(\g_1,\g_2)+H^{(3)}(\g_1,\g_2)\,.
\end{equation}
The gluon on-shell limit can be investigated by examining the residues
of the poles when $\g \rightarrow 0$.  We define the function $h$
\begin{equation}
h(\g_1,\g_2)= \g_1 \g_2  \, H_{0} (\g_1,\g_2)\,. 
\end{equation}
The result when we take one leg on shell is known, $\g_2 \rightarrow
0$ (see ref.~\cite{Ellis:1990hw})
\begin{eqnarray} 
\label{onelegonshell}
h(\g_1,0) &=& \frac{\pi \alpha_s^2}{\V} B(1+\g_1,1-\g_1)
B(1-\g_1,1-\g_1) \times
\nn \\&& 
\times\Biggl[ \N \frac{2 (2-\g_1)}{(3-2 \g_1) (5-2 \g_1)} -\frac{1}{ 2
\N}\frac{(7-5 \g_1)}{3 (3-2 \g_1)} \Biggr]\,.
\end{eqnarray}
Note that the colour suppressed piece of this expression gives the
contribution in an abelian theory. It is thus related to the impact
factor for the direct photoproduction of heavy quarks: in the notation
of eq.~(\ref{Jjdef}), $j(\g) = -e_Q^2\frac{\alpha_{em}}{\as}\, 2 \V\N \,
h^{(1)}(\g,0)$.  We can also check the zeroth moment of the normal
cross section when both gluons are on shell,
cf.~eq.~(\ref{NDEresults})
\begin{eqnarray} 
\label{twolegsonshell}
h(0,0) &=& \frac{\pi \alpha_s^2}{\V}  
 \Biggl[ \N \frac{4}{15}-\frac{1}{ \N}\frac{7}{18} \Biggr]\,.
\end{eqnarray}

As a final check we have evaluated eq.~(\ref{Hequation}) for 
$H_0$ numerically and find good agreement with 
eq.~(\ref{Hresult}) in the range $0<\g_1,\g_2<{1}/{2}$.

The triple pole in eq.~(\ref{Hthree}) may be of some phenomenological
importance. It is easy to evaluate $H^{(3)}$ for arbitrary $\o$ leading to a
contribution, which for $\g_1,\g_2 \sim {1}/{2}$ may be written as,
\begin{equation}
H^{(3)}_\o(\g_1,\g_2)\sim  \frac{\pi \alpha_s^2}{\V}
\f{\N}{6} \frac{1} {\g_1 \g_2 (1-\g_1-\g_2)(1-\g_1-\g_2+\o)^2}\,.
\end{equation}

\section{The $K$ functions} \label{appb}

Using the saddle point method in the vicinity of a simple pole we
have to evaluate an integral of the form
\begin{eqnarray}
K_1(y) &=&\int_{-\infty}^{\infty} dz \f{b}{(b^2+z^2)}
\exp\left(-\f{z^2}{c^2}\right) 
\nn \\
&=& \pi \exp(y^2) \left[1- \mbox{Erf}(y) \right] \equiv
\pi \exp(y^2) \mbox{Erfc}(y)\,,
\end{eqnarray}
with $y=b/c$ and $b,c>0$. Similiarly, for integrals in the
neighbourhood of a double pole it is useful to define
\begin{eqnarray}
K_2(y) &=& c \int_{-\infty}^{\infty} dz \f{b^2-z^2}{(b^2+z^2)^2}
\exp \left(-\f{z^2}{c^2}\right) 
\nn \\
&\equiv& 2 \sqrt{\pi} -2 y K_1(y)\,,
\end{eqnarray}
and near a triple pole
\begin{eqnarray}
K_3(y) &=& c^2 b \int_{-\infty}^{\infty} dz \f{b^2-3z^2}{(b^2+z^2)^3}
\exp \left(-\f{z^2}{c^2}\right) 
\nn \\
&\equiv& K_1(y) - y K_2(y)\,.
\end{eqnarray}
\FIGURE[t]{\epsfig{file=kplot.ps,width=.7\textwidth}
\caption{Plot of the functions $K_1(y)$, $K_2(y)$ and
$K_3(y)$.} \label{k}}

The functions $K_1,K_2,K_3$ have the following expansions: for
small $y$
\begin{eqnarray}
K_1(y) &=& \pi \left[1-\f{2 y }{\sqrt\pi} +y^2 + O(y^3)\right],
\nn\\
K_2(y) &=& 2 \sqrt{\pi} \left[1-\sqrt{\pi}y +2 y^2 + O(y^3)
\right],
\nn\\
K_3(y) &=&\pi \left[1-\f{4 y }{\sqrt\pi} +3y^2 + O(y^3)\right],
\end{eqnarray}
while for large $y$
\begin{eqnarray}
K_1(y)&=& \f{\sqrt{\pi}}{y} \left[1-\f{1}{2 y^2}+\f{3}{4 y^4}+
O\left(\f{1}{y^6}\right)\right],
\nn\\
K_2(y)&=& \f{\sqrt{\pi}}{y^2} \left[1-\f{3}{2 y^2}+
O\left(\f{1}{y^4} \right) \right],
\nn\\
K_3(y)&=& \f{\sqrt{\pi}}{y^3} \left[1+\f{3}{4 y^2}+ O
\left(\f{1}{y^4}\right)\right].
\end{eqnarray}
A plot of the functions $K_1$, $K_2$ and $K_3$ is shown in
figure~\ref{k}.


\begin{thebibliography}{99}

\bibitem{ref:DO_bx}
{\sc D0} collaboration,  \emph{Inclusive $\mu$ and b
  quark production cross-sections in $p \bar p$ collisions at $\sqrt{s} =
  1.8$\,TeV}, \prl{74}{1995}{3548}.

\bibitem{ref:DO_dim}
DO collaboration, \emph{The $b\bar b$ production cross-section and
angular correlations in $p\bar p$ collisions at $\sqrt{s} = 1.8$\,TeV}
\plb{487}{2000}{264} [\hepex{9905024}].


\bibitem{ref:CDF_bx}
{\sc CDF} collaboration,  \emph{Measurement of the
  bottom quark production cross-section using semileptonic decay
  electrons in $p\bar p$ collisions at $\sqrt{s} = 1.8$\,TeV},
  \prl{71}{1993}{500};
\emph{Measurement of bottom
  quark production in 1.8\,TeV $p\bar p$ collisions using semileptonic
  decay muons}, \prl{71}{1993}{2396};\\
\emph{Inclusive $\chi_c$ and b-quark production in $\bar p p$ collisions
  at $\sqrt{s} = 1.8$\,TeV}, \prl{71}{1993}{2537};\\
\emph{Measurement of the B meson differential cross-section, $d \sigma/d
  p_T$, in $p\bar p$ collisions at $\sqrt{s} = 1.8$\,TeV},
  \prl{75}{1995}{1451} [\hepex{9503013}].

\bibitem{Abbott:2000wu}
{\sc D0} collaboration, \emph{Small angle muon and
  bottom quark production in $p\bar p$ collisions at $\sqrt{s} =
  1.8$\,TeV}, \prl{84}{2000}{5478} [\hepex{9907029}].

\bibitem{Nason:1988xz}
P.~Nason, S.~Dawson and R.K. Ellis, \emph{The total cross-section for
  the production of heavy quarks in hadronic collisions},
  \npb{303}{1988}{607}.

\bibitem{Mangano:1992jk}
M.L. Mangano, P.~Nason and G.~Ridolfi, \emph{Heavy quark correlations
  in hadron collisions at next-to-leading order},
  \npb{373}{1992}{295}.

\bibitem{Beenakker:1991ma}
W.~Beenakker, W.L. van Neerven, R.~Meng, G.A. Schuler and J.~Smith,
  \emph{QCD corrections to heavy quark production in hadron-hadron
  collisions}, \npb{351}{1991}{507}.

\bibitem{Adloff:1999nr}
H1 collaboration,
\emph{Measurement of open beauty production at HERA},
\plb{467}{1999}{156}
[\hepex{9909029}].

\bibitem{Breitweg:2000nz}
ZEUS collaboration, \emph{Measurement of open beauty production in
photoproduction at HERA}, \epjc{18}{2001}{625} [\hepex{0011081}].

\bibitem{Acciarri:2000kd}
{\sc L3} collaboration, \emph{Measurements of the
  cross sections for open charm and beauty production in $\gamma \gamma$
  collisions at $\sqrt{s} = 189$\,GeV -- $202$\,GeV}, \plb{503}{2001}{10}
  [\hepex{0011070}].

\bibitem{Csilling:2000xk}
{\sc OPAL} collaboration, \emph{Charm and bottom
  production in two-photon collisions with OPAL}, \hepex{0010060}.

\bibitem{Frixione:1997nh}
S.~Frixione and M.L. Mangano, \emph{Heavy-quark jets in hadronic
  collisions}, \npb{483}{1997}{321} [\hepph{9605270}].

\bibitem{Abbott:2000iv}
{\sc D0} collaboration,  \emph{Cross section for B
  jet production in $\bar p p$ collisions at $\sqrt{s} = 1.8$\,TeV},
  \prl{85}{2000}{5068} [\hepex{0008021}].

\bibitem{Bonciani:1998vc}
R.~Bonciani, S.~Catani, M.L. Mangano and P.~Nason, \emph{NLL
  resummation of the heavy-quark hadroproduction cross-section},
  \npb{529}{1998}{424} [\hepph{9801375}].

\bibitem{Ellis:1990hw}
R.K. Ellis and D.A. Ross, \emph{The coupling of the QCD pomeron in
  various semihard processes}, \npb{345}{1990}{79}.

\bibitem{Catani:1990xk}
S.~Catani, M.~Ciafaloni and F.~Hautmann, \emph{Gluon contributions to
  small x heavy flavor production}, \plb{242}{1990}{97}.

\bibitem{Collins:1991ty}
J.C. Collins and R.K. Ellis, \emph{Heavy quark production in very
  high-energy hadron collisions}, \npb{360}{1991}{3}.

\bibitem{Catani:1991eg}
S.~Catani, M.~Ciafaloni and F.~Hautmann, \emph{High-energy
  factorization and small x heavy flavor production},
  \npb{366}{1991}{135}.

\bibitem{Catani:1992zc}
S.~Catani, M.~Ciafaloni and F.~Hautmann,
\emph{Production of heavy flavors at high-energies},
CERN-TH-6398-92, presented at the Workshop on Physics at
HERA, Hamburg, Germany, Oct 29-30, 1991.

\bibitem{Camici:1996st}
G.~Camici and M.~Ciafaloni, \emph{Non-abelian $q \bar q$ contributions
  to small-x anomalous dimensions}, \plb{386}{1996}{341}
  [\hepph{9606427}].

\bibitem{Camici:1997ta}
G.~Camici and M.~Ciafaloni, \emph{k-factorization and small-x
  anomalous dimensions}, \npb{496}{1997}{305} [\hepph{9701303}].

\bibitem{Hagler:2000dd}
P.~Hagler, R.~Kirschner, A.~Schafer, L.~Szymanowski and O.~Teryaev,
  \emph{Heavy quark hadroproduction as a test of the effective BFKL $q
  \bar q$ vertex}, \prd{62}{2000}{071502} [\hepph{0002077}].

\bibitem{Ryskin:2000wz}
M.G. Ryskin, Y.M. Shabelski and A.G. Shuvaev, \emph{Heavy quark
  production in hadron collisions}, \hepph{0011111}.

\bibitem{Ball:1994du}
R.D. Ball and S.~Forte, \emph{Double asymptotic scaling at HERA},
\plb{335}{1994}{77} [\hepph{9405320}];\\
\emph{A direct test of perturbative QCD at small x},
  \plb{336}{1994}{77} [\hepph{9406385}].

\bibitem{Ball:1995vc}
R.D. Ball and S.~Forte, \emph{Summation of leading logarithms at small
  x}, \plb{351}{1995}{313} [\hepph{9501231}].

\bibitem{Ellis:1995gv}
R.K. Ellis, F.~Hautmann and B.R. Webber, \emph{QCD scaling violation
  at small x}, \plb{348}{1995}{582} [\hepph{9501307}].

\bibitem{Forte:1996xv}
S.~Forte and R.D. Ball, \emph{Double scaling violations},
  \hepph{9607291}.

\bibitem{Bojak:1997nr}
I.~Bojak and M.~Ernst, \emph{Small x resummations confronted with
  $F_2(x,Q^2)$ data}, \plb{397}{1997}{296} [\hepph{9609378}];\\
\emph{Limitations of small x resummation methods from $F_2$ data},
  \npb{508}{1997}{731} [\hepph{9702282}].

\bibitem{Blumlein:1998em}
J.~Blumlein and A.~Vogt, \emph{The evolution of unpolarized singlet
  structure functions at small x}, \prd{58}{1998}{014020}
  [\hepph{9712546}].

\bibitem{Fadin:1998py}
V.S. Fadin and L.N. Lipatov, \emph{Bfkl pomeron in the next-to-leading
  approximation}, \plb{429}{1998}{127} [\hepph{9802290}].

\bibitem{Ross:1998xw}
D.A. Ross, \emph{The effect of higher order corrections to the BFKL
  equation on the perturbative pomeron}, \plb{431}{1998}{161}
  [\hepph{9804332}].

\bibitem{Ball:1998be}
R.D. Ball and S.~Forte, \emph{Corrections at small x},
  \hepph{9805315}.

\bibitem{Kovchegov:1998ae}
Y.V. Kovchegov and A.H. Mueller, \emph{Running coupling effects in
  BFKL evolution}, \plb{439}{1998}{428} [\hepph{9805208}].

\bibitem{Armesto:1998gt}
N.~Armesto, J.~Bartels and M.A. Braun, \emph{On the 2nd order
  corrections to the hard pomeron and the running coupling},
  \plb{442}{1998}{459} [\hepph{9808340}].

\bibitem{Salam:1998tj}
G.P. Salam, \emph{A resummation of large sub-leading corrections at
  small x}, \jhep{07}{1998}{019} [\hepph{9806482}].

\bibitem{Ball:1999sh}
R.D. Ball and S.~Forte, \emph{The small x behaviour of
  Altarelli-Parisi splitting functions}, \plb{465}{1999}{271}
  [\hepph{9906222}].

\bibitem{Ciafaloni:1999yw}
M.~Ciafaloni, D.~Colferai and G.P. Salam, \emph{Renormalization group
  improved small-x equation}, \prd{60}{1999}{114036}
  [\hepph{9905566}].

\bibitem{Altarelli:2000vw}
G.~Altarelli, R.D. Ball and S.~Forte, \emph{Resummation of singlet
  parton evolution at small x}, \npb{575}{2000}{313}
  [\hepph{9911273}].

\bibitem{Altarelli:2000mh}
G.~Altarelli, R.D. Ball and S.~Forte, \emph{Small-x resummation and
  hera structure function data}, \npb{599}{2001}{383}
  [\hepph{0011270}].

\bibitem{Ellis:1989sb}
R.K. Ellis and P.~Nason, \emph{QCD radiative corrections to the
  photoproduction of heavy quarks}, \npb{312}{1989}{551}.

\bibitem{Catani:1996ze}
S.~Catani, \emph{Comment on quarks and gluons at small x and the SDIS
  factorization scheme}, \zpc{70}{1996}{263} [\hepph{9506357}];
\emph{Quarks and gluons at small x and scaling violation of $F_2$},
  \hepph{9506348}.

\bibitem{Ball:1995tn}
R.D. Ball and S.~Forte, \emph{Momentum conservation at small x},
\plb{359}{1995}{362} [\hepph{9507321}].

\bibitem{Catani:1994sq}
S.~Catani and F.~Hautmann, \emph{High-energy factorization and small x
  deep inelastic scattering beyond leading order},
  \npb{427}{1994}{475} [\hepph{9405388}].

\bibitem{DeRujula:1974rf}
A.D. Rujula, S.L.~Glashow, H.D.~Politzer, S.B.~Treiman, F.~Wilczek
and A.~Zee, \emph{Possible nonregge behavior of electroproduction
structure functions}, \prd{10}{1974}{1649}.

\end{thebibliography}
\end{document}